\documentclass[12pt]{article} 

\input epsf.tex

\newcommand{\be}{\begin{equation}}
\newcommand{\ee}{\end{equation}}
\newcommand{\bea}{\begin{eqnarray}}
\newcommand{\eea}{\end{eqnarray}}

\newcommand{\gs}{\ensuremath{g_s}} 
\newcommand{\ap}{\ensuremath{\alpha'}} 
\newcommand{\ls}{\ensuremath{l_s}} 

\newcommand{\hepth}[1]
{\href{http://xxx.lanl.gov/abs/hep-th/#1}{{\tt hep-th/#1}}}

\def\eps{\epsilon}

\def\p{\partial}

\newcommand{\Rea}{{\mathrm{Re}}}
\newcommand{\Ima}{{\mathrm{Im}}}
\newcommand{\cO}{{\mathcal{O}}}
\newcommand{\cN}{{\mathcal{N}}}

\newcommand{\bS}{{\mathbf{S}}}

\newcommand{\Gs}{\ensuremath{G_s}} 
\newcommand{\Go}{\ensuremath{G_o^2}} 
\newcommand{\ape}{\ensuremath{\alpha'_{e}}} 
\newcommand{\Ls}{\ensuremath{L_s}} 
\newcommand{\tR}{\ensuremath{\tilde{R}}}
\newcommand{\tRm}{\ensuremath{\tilde{R}_{-}}}

\newcommand{\Pp}{\ensuremath{P'_{+}}}
\newcommand{\Pm}{\ensuremath{P'_{-}}}

\begin{document}

\begin{titlepage}

\begin{flushright}
UUITP-08/00\\
USITP-00-15\\  
hep-th/0009182
\end{flushright}

\vspace{1cm}

\begin{center}
{\huge\bf IIA/B, Wound and Wrapped}
\end{center}
\vspace{5mm}

\begin{center}

{\large Ulf H.\ Danielsson,$^{\scriptstyle 1}$
Alberto G\"uijosa,$^{\scriptstyle 2}$
and
Mart\'\i n Kruczenski$^{\scriptstyle 1}$} \\

\vspace{5mm}

$^{1}$ Institutionen f\"or Teoretisk Fysik, Box 803, SE-751 08
Uppsala, Sweden

\vspace{3mm}

$^{2}$ Institute of Theoretical Physics, Box 6730, SE-113 85
Stockholm, Sweden

\vspace{5mm}

{\tt
ulf@teorfys.uu.se, alberto@physto.se, \\
martin.kruczenski@teorfys.uu.se \\
}

\end{center}

\vspace{5mm}

\begin{center}
{\large \bf Abstract}
\end{center}
\noindent

We examine the T-duality relation between $1+1$
NCOS and the DLCQ limit of type IIA string theory. We show that, as
long as there is a compact dimension,
one can meaningfully define an `NCOS' limit of IIB/A string theory
even in the absence of D-branes
(and even if there is no $B$-field). This yields a theory of closed
strings with strictly positive winding, which is T-dual to DLCQ IIA/B
without any D-branes. We call this the
Type IIB/A Wound String Theory.
The existence of decoupled sectors can be seen
directly from the energy spectrum, and mirrors that of the DLCQ theory.
It becomes clear then that all of the
different $p+1$ NCOS theories are simply different states of this
single Wound IIA/B theory which contain D-branes.
We study some of the properties of this theory. In
particular, we show that upon toroidal compactification, Wound string
theory is U-dual to various
Wrapped Brane theories which contain OM theory
and the OD$p$ theories as special states.

\vfill
\begin{flushleft}
September 2000
\end{flushleft}
\end{titlepage}
\newpage

\section{Introduction}

During the past couple of years there has been a lot of progress in the
understanding of D-branes in the presence of magnetic fields and
how these can be used to study Yang-Mills field theories with space/space
non-commutativity \cite{cds,dh,sw}. 
Recently, as a generalization of this, the physics of
D-branes with near-critical electric fields on their world-volumes has
also attracted considerable attention \cite{sst2,ncos,om}.  Such theories
have a peculiar space/time non-commutativity\footnote{See \cite{agm}
for a discussion of theories with lightlike non-commutativity.} 
which in principle threatens
their consistency \cite{sst1,br,gm}. 
However, as has been argued in the above
cited works, the resulting theory is not a field theory, 
but a theory of open strings with a string scale of the same 
magnitude as the non-commutativity parameter. 
Contrary to the case of magnetic fields and
space/space non-commutativity, then,
the space/time non-commutativity and the
stringiness cannot be disentangled from one another. 
The resulting theories are
known as $(p+1)$-dimensional
non-commutative open string theories (NCOS)  \cite{sst2,ncos,om}.  
The $3+1$  NCOS theory turns out to be
S-dual to a theory with ordinary space/space non-commutativity,
namely, $3+1$ NCYM \cite{ncos} (see also \cite{grs}).

Several successful calculations have been performed in NCOS theories.
Examples include scattering amplitudes in \cite{ncos,km,hk}, 
and supergravity duals and finite
temperature physics in \cite{harmark1,sahakian,ggkrw} 
(see also \cite{gkp}-\cite{gremm}).
NCOS theories have the usual rules for
computing scattering amplitudes, except for the appearance of Moyal phases
which depend on the ordering
of the open string vertex operators \cite{sst2,ncos,km}, and 
are thus the source of non-commutativity. 
The closed string sector can be seen to
decouple leaving a theory which contains only open strings
\cite{sst2,ncos}, unless, as Klebanov and
Maldacena have pointed out \cite{km}, one compactifies the theory on a
circle along the direction of the electric field. 
One then finds that closed strings with strictly positive
winding number have a
finite energy in the NCOS limit,
and interact in a non-trivial way with the brane. 

It has also been noted in \cite{sst2,km,kleb00} that upon
compactification along the direction of the electric field
NCOS theories are T-dual  to
the discrete light-cone quantization (DLCQ) of Type II theories. 
DLCQ of M-theory and string theory has been the subject of intensive research
in the Matrix theory context \cite{bfss,susskind}.
As explained by Seiberg \cite{seiberg} (see also \cite{sen}), 
DLCQ can be defined by
considering the theory on a spatial circle of vanishing size and then
boosting along the direction of the circle to get a (finite-size) almost
lightlike circle. The T-duality between the IIA/B NCOS-theory and the DLCQ
of IIB/A arises because a D$p$-brane with a near-critical electric 
field T-dualizes to a D$(p-1)$-brane moving at the speed close to that
of light \cite{km,sst2}. This is, in
fact, precisely the condition for the D$(p-1)$-brane to remain in the 
DLCQ spectrum after the boost: if the D-brane were instead at rest 
it would acquire an infinite energy. A more
detailed exploration of this T-duality relation between the NCOS and DLCQ 
limits is the central subject of the present paper.

We begin in Section \ref{dlcqsec} with a brief review of the NCOS 
limit, focusing for concreteness on the $(1+1)$-dimensional case. 
We then proceed to exhibit the explicit mapping between the
NCOS and DLCQ descriptions. In the process, we encounter a small puzzle.
Starting with a brane carrying an electric field,
T-duality will transform the
electric field into a velocity. By performing a boost one can then put the
brane at rest. The worldsheet fields will then obey the usual
Neumann/Dirichlet boundary conditions, and one would not expect any Moyal
phases associated with non-commutativity even in the DLCQ limit--- the 
T-dual image of the NCOS limit. The resolution of this is to remember that
T-duality requires compactification on a circle, and converts momentum 
into winding number.  The left- and right-moving momenta are then 
different. It will be shown in Section \ref{dlcqsec}
that combining this with the fact that the
circle is boosted, one indeed reproduces the expected Moyal phases,
even though they now have a different origin.

A related puzzle is the following.
As explained in \cite{km}, the closed string spectrum of the 
compactified $1+1$ NCOS theory is the one which results from
the presence of a near-critical $B_{01}$-field.  
As can be seen from the formulas in that paper,
such a $B$-field in fact modifies
the mass-shell condition for closed strings in a
way that is very similar to the light-cone energy-momentum relation.
At first sight this is somewhat confusing, since the $B$-field 
can in this case (as opposed to the usual case of a $B_{ij}$-field
on a two-torus) be gauged
away from the bulk and into the D-brane, where it cannot have any 
effect on the closed string spectrum. As we
will discuss in detail in Section \ref{clstrsec}, 
the key point is that the $B$-field is merely an artifact that plays
the same role as the boost in the Seiberg procedure 
\cite{seiberg,sen}. A boost does not
change the physics, but in the case of \cite{seiberg,sen}
maps the variables to simpler ones. In particular, it
subtracts an infinite contribution from the energy.

It follows from this observation that it should be possible to deduce 
the decoupling of closed strings with non-positive winding directly
in the gauge where it is $F_{01}$ and not $B_{01}$ that becomes 
critical, and in Section \ref{clstrsec} we verify that this is indeed 
the case. It becomes apparent then that this decoupling has nothing to 
do with the presence of the D-branes or background fields, and that, 
as long as there is a compact direction, one can 
meaningfully define an `NCOS' limit of the full IIA/B string theory 
(with or without branes). We study the resulting ten-dimensional
theory in Section \ref{ncossec}, and we discover that its defining 
property is the fact that all objects in it must carry strictly 
positive F-string winding.  It is thus natural to call this the (IIA 
or IIB) Wound String theory. It is T-dual to DLCQ (IIB or IIA) string 
theory, and the corresponding
decoupling arguments are mirror images of one another.
The various $p+1$ NCOS 
theories are then different states in this single unifying theory--- 
namely, those states that contain a D$p$-brane wrapping the 
compact direction. 
We emphasize that the Wound theory is well-defined and 
non-trivial even in the absence of such branes.

In Section \ref{ncosontorussec} we examine the Wound IIA/B theories 
compactified on transverse tori. Through diverse dualities, we 
are driven to define various 
Wrapped $p$-brane theories which are the natural generalizations of
the Wound string idea: these theories are obtained as limits of 
string/M-theory with (at least) $p$ compact directions, and all 
objects in them must carry strictly positive $p$-brane wrapping 
number on this $p$-torus.  Just like the various NCOS theories
are different classes of states in a single Wound string 
theory, the recently discovered OM theory 
\cite{om,bbss2} (see also \cite{bbss1}) and OD$p$ 
theories \cite{harmark2,om} are 
specific classes of states in a broader framework: as it will be shown 
in Section \ref{ncosontorussec}, OM theory is
the Wrapped M2-brane theory in the presence of M5-branes which wrap 
the `Wrapped' directions, whereas 
each of the OD$p$ theories is understood to be
the Wrapped D$p$-brane theory in the presence of NS5-branes.
Section \ref{ncosontorussec} includes a discussion of 
the relation of these theories to the known S-duals for NCOS 
theories \cite{ncos,om,kt} and the theories encountered in Seiberg's 
derivation of the Matrix description for DLCQ 
IIA/M-theory \cite{seiberg,sen}.  

We conclude in Section \ref{conclusions}, where we in particular
emphasize the distinction between the Matrix and Wrapped points
of view.

\section{1+1 NCOS vs. DLCQ IIA} \label{dlcqsec}

\subsection{Review of the NCOS limit} \label{ncosrevsec}

In this section we will
review the NCOS limit defined in~\cite{sst2,ncos,om} 
specializing
to the $(1+1)$-dimensional case; the higher-dimensional
cases will be discussed in Section \ref{ncosontorussec}. 
Consider then an $(N,1)$ string, i.e., a bound state of a D-string and $N$
fundamental strings. We take the string to lie along the $x^{1}$ direction,
and the background (string frame) metric to be flat and diagonal,
\begin{equation} \label{metric}
g_{ab}=\eta _{ab},\qquad g_{ij}=h\delta _{ij},\qquad a,b=0,1,\qquad
i,j=2,\ldots ,9~. \label{g}
\end{equation}
As indicated, we have split the ten spacetime directions into those parallel
and perpendicular to the string, $\mu =(a,i)$.

The $N$ fundamental strings in the bound state are represented as $N$ units
of electric flux. The relation between the electric field and $N$ follows
from the Dirac-Born-Infeld action (see, e.g., \cite{ck}),
\be \label{quant}
\frac{2\pi\ap F_{01}}{\sqrt{1-(2\pi\ap F_{01})^{2}}}=N\gs\quad
\Longrightarrow\quad
E = \frac{N\gs}{\sqrt{(N\gs)^{2}+1}}~,
\ee
where $E\equiv 2\pi\ensuremath{\alpha'} F_{01}$
is the electric field measured in units
of its critical value.

The net effect of the electric field on open string dynamics can be
summarized by introducing the effective string metric, coupling constant,
and non-commutativity parameter \cite{sw,sst2}
\be \label{swmap}
G_{ab}=(1-E^{2})\eta_{ab}, \quad G_{ij}=h\delta_{ij}; \qquad
G_{o}^{2}=\gs\sqrt{1-E^{2}}; \qquad \theta={2\pi\ap}{E\over 1-E^{2}}~.
\ee

Let $E=1-\epsilon/2$. The non-commutative open string (NCOS) limit \cite
{sst2,ncos,om} is a near-critical limit
\be \label{gmmsslim}
\eps\to 0, \quad\mbox{with}\quad \ap=\ap_{e}\eps\to 0, \quad h=\eps\to 0;
\quad \ap_{e},N \quad\mbox{fixed}~.
\ee
Notice that an effective Regge slope, $\ap_{e}$, has been introduced;
it is this slope which is held fixed in the limit. 

It follows from (\ref{quant}) and
(\ref{swmap}) that in the NCOS limit\footnote{The near-critical
limit for the $(N,1)$ string, but taking $\gs$ fixed and $N$ large, as
opposed to $N$ fixed and $\gs$ large, was
analyzed in \cite{gkp,verlinde,sst2}. The physics of the two limits is
in some respects similar.}
\bea \label{gmmssout}
\gs&=& {1\over N\sqrt{\eps}}\to\infty~, \\
\label{gmmssout2}
\Go&=& {1\over N}~, \\
\label{gmmssout3}
G_{\mu\nu}&=& \eps\eta_{\mu\nu}~, \\
\label{gmmssout4}
\theta&=& 2\pi\ap_{e}~.
\eea
The end result is a string theory of
open strings
with tension set by \ape, coupling constant \Go, and non-commuting
$x^{0},x^{1}$ directions, $[x^{0},x^{1}]\sim i\theta$
\cite{sst2,ncos,om}.
The scaling of the parameter $h$
in the transverse
closed string metric (\ref{g}) expresses the
rescaling of distances which is needed for the limit to produce
a full-fledged string theory. As one approaches criticality, the
strings in effect become tensionless only in the direction of the
electric field, so one must zoom-in on a small transverse region
(or equivalently, scale down the longitudinal direction)
in order for the
transverse oscillations of the strings to remain in the theory.
Were it not for this rescaling, the strings would behave as rigid rods.
Notice also that the fact that $G_{\mu\nu}$ in (\ref{gmmssout3})
vanishes as $\eps\to 0$ is
compensated in the open-string
mass-shell condition by
the rescaling of \ap, so the metric of the NCOS theory
is in effect $\eta_{\mu\nu}$.

Next, we compactify the theory along the direction of the electric
field (and the string), identifying $x^{1}\simeq x^{1}+2\pi R$,
and regarding $R$ as fixed in the NCOS limit.
It has been shown
in \cite{km} that wound closed strings are then present in the
spectrum of the
theory (these become infinitely massive in the infinite volume
limit, of course). The intuitive picture is as follows. The near-critical
electric field forces the ends of the open strings to move away from
each other along the positive $x^{1}$ direction. In the non-compact
case, it would thus be energetically forbidden (in the strict NCOS limit)
for the ends to join together to form a closed
string. If the $x^{1}$ direction is compact, on the other hand, the
ends can meet after having moved around the circle a certain number of
times, thus giving rise to closed strings with strictly positive
winding number \cite{km}, $w>0$. 

Despite this intuitive picture, it is surprising that the presence
of the electric field could have an effect on the \emph{closed} 
strings, since they are supposed to be electrically neutral.
Klebanov and Maldacena \cite{km} circumvented this difficulty by
gauging the electric field $E$ on the brane into a spacetime $B$-field
--- it is then easy to see the change in the
closed string energy spectrum as $B$ approaches criticality.
In Section \ref{clstrsec}
we will discuss this in detail, showing that
the gauge transformation
that shifts $E$ into $B$ is T-dual to a
boost. We will also explain how the decoupling of closed strings with
$w\leq 0$ can be seen directly in
the $(E\neq 0, B=0)$ gauge,
and from this it will become clear that the same
decoupling will take place in the absence of any D-branes---
even if there is no $B$-field!

\subsection{Longitudinal T-duality and Matrix String Theory}
\label{ncostsec}

We are now interested in studying the effect of T-duality on the
NCOS theory. It was pointed out in~\cite{sst2,km,kleb00} that
this should yield the DLCQ IIA
string theory. Here we try to make this notion
more precise.
For this we will describe the
theory in terms of the parent IIB theory, with a single D1-brane
and a near-critical electric field $E=1-\eps/2$, $\eps\to 0$.
T-dualizing along the compact direction, we obtain an
equivalent description of the system in terms of a D0-brane moving 
with speed close to that of light, 
\be \label{v}
v=\p_{0}X^{1}=1-\eps/2,
\ee
in a IIA theory whose parameters follow from
the rules of T-duality:
\be \label{T}
\left\{
\begin{array}{lcl}
g_s &=& \Go/\sqrt{\epsilon} \\
R_{\perp} &\sim& \sqrt{\epsilon} \\
R_1 &\sim& 1 \\
\ap &=& \ape \epsilon
\end{array}
\right. \qquad
{\mbox{T}_{1} \atop \longleftrightarrow} \qquad
\left\{
\begin{array}{lcl}
\tilde{g}_s &=& g_s l_s/R_1 = \Go \sqrt{\ape}/{R_1} \\
\tilde{R}_\perp &\sim& \sqrt{\epsilon} \\
\tilde{R}_1 &=& \ap/R_1 = \ape \epsilon/{R_1} \\
\ap &=& \ape \epsilon
\end{array}
\right.
\ee
Here $R_\perp$ denotes any (proper) length 
in the transverse (non-compact) directions. The scaling 
$R_{\perp}\sim\sqrt{\epsilon}$ results from 
the transverse metric (\ref{metric}).
The $N$ units of fundamental string winding on the D1-brane
become $N$ units of 
Kaluza-Klein momentum for the D0-brane, $P_{1}=N/\tR\to\infty$.

The scaling in the right-hand side of (\ref{T})
is precisely the one that defines 
the DLCQ limit of IIA string theory,
in the sense of ~\cite{seiberg,sen}.  Following Seiberg, 
we rescale all lengths in the theory
by a factor of $\sqrt{\epsilon}$, to work in units in which the 
string length is finite,
\be \label{lst}
\tilde{\ls}=\sqrt{\ape}~.
\ee
The change of units makes all quantities finite except $\tilde{R}_1$,
which becomes $\tilde{R}_1=\sqrt{\epsilon}\ape/R_1$.
We next carry out a large boost along direction $1$, 
with velocity $\beta=1-\epsilon/2$, 
to a frame F$'$ where 
the circle lies (almost)
along $x'^{-}\equiv (x'^{0}-x'^{1})/2$ and has a finite
radius\footnote{Boosting 
the system by an additional finite amount just rescales this null 
radius. We have chosen the boost 
parameter to equal the velocity (\ref{v}),
so the D0-brane
will in fact be at rest along the $x'^{1}$ direction in the frame
F$'$. There is a free parameter  
describing the `finite portion' of the boost, and (\ref{trm})
reflects our specific choice for this parameter, which
translates into a specific relation between the NCOS and DLCQ
reference frames F and F$'$.}
\be \label{trm}
\tRm ={\tR\over\eps}={\ape\over R_{1}}~.
\ee

We have thus shown explicitly that $(1+1)$-dimensional NCOS with
parameters $\Go=1/N,\ape$, compactified on a circle of radius $R$, is
T-dual to DLCQ IIA string theory
with coupling constant
\be \label{couplingcst}
\tilde{\gs}=\Go {\sqrt{\ape}\over R_{1}}~,
\ee
string length  (\ref{lst}), and null radius (\ref{trm}), 
in the presence of a single D0-brane, and with
total longitudinal momentum $\Pm=N/\tRm$.
Notice that (\ref{lst}),
(\ref{trm}) and  (\ref{couplingcst}) are in fact just the
naive T-duality relations,
but they use \ape\ instead of \ap, and relate a spatial
compactification to a lightlike one. This relation
evidently extends to the case
with an arbitrary number, $K$, of D1-branes:
$1+1$ NCOS$(K)$
with parameters $\Go=K/N,\ape$, on a circle of radius $R$,
is T-dual to DLCQ IIA
with parameters $\tilde{\gs}=\Go\sqrt{\ape}/R$ and (\ref{lst}), on a
circle of radius (\ref{trm}),
in the presence of $K$ D0-branes, and with $N$ units of
longitudinal momentum.

Notice that we are using the term `T-duality' 
in a slightly extended sense, 
since the mapping that relates the two 
theories is the composition of
ordinary T-duality, a change of units, 
and a boost. The change of units can of course
be avoided by presenting the NCOS 
theory directly in the new units: this would entail keeping $\ls$ and 
the transverse metric parameter $h$ fixed, while 
scaling the parallel components of the metric
according to $g_{ab}=\eps^{-1}\eta_{ab}$. As shown in \cite{sst2}, 
this results in an NCOS theory with parameters $\Go=1/N$, $\ape=\ap$, 
$\theta=2\pi\ap$,  $G_{\mu\nu}=\eta_{\mu\nu}$. 
In Section \ref{boostsec}
we will see that the boost can also be avoided, 
by working in the gauge where it is $B$ and not $E$ that becomes 
critical. 

We note in passing that the above
DLCQ IIA string theory (after the change of units)
can be equivalently formulated
as M-theory with eleven-dimensional Planck length
\be \label{tlP}
\tilde{l}_{P}=\tilde{\gs}^{1/3}\tilde{\ls}=\left({\alpha'^{2}_{e}\over N
R}\right)^{1/3}~,
\ee
compactified on a null circle of radius (\ref{trm}), and
a transverse spatial circle of radius
\be \label{rt10}
\tilde{R}_{10}=\tilde{\gs}\tilde{\ls}={\ape\over N R}~,
\ee
with $N$ units of longitudinal momentum, and one unit of transverse
momentum.

DLCQ IIA string theory is believed to admit a non-perturbative
description in terms of Matrix string theory  
(i.e., $\cN=8$ super-Yang-Mills in $1+1$ dimensions)
\cite{motl,bs,dvv},
and the series of steps we have followed
above is in fact exactly the reverse
of the sequence of dualities used to justify \cite{seiberg}
the Matrix string theory conjecture. The
situation is summarized in Fig.~1.

\begin{figure}[htb]
\begin{center}
\begin{picture}(150,45) {\small
\put(0,40){\begin{minipage}[t]{4cm} \begin{center}
1+1 NCOS (IIB) \\
N F-strings \\
K D1-branes
\end{center}\end{minipage}}
\put(0,10){\begin{minipage}[t]{4cm} \begin{center}
1+1 SYM ($\overline{\mbox{IIB}})$\\
$U(N)$ gauge group \\
K units of $E$-flux
\end{center}\end{minipage}}
\put(55,40){\begin{minipage}[t]{4cm} \begin{center}
IIA on small $\mathbf{S}^{1}$\\
N units of $P_{1}$ \\
K D0-branes
\end{center}\end{minipage}}
\put(55,10){\begin{minipage}[t]{4cm} \begin{center}
$\overline{\mbox{IIA}}$ on small $\mathbf{S}^{1}$\\
N D0-branes\\
K units of $P_{1}$
\end{center}\end{minipage}}
\put(110,40){\begin{minipage}[t]{4cm} \begin{center}
DLCQ IIA \\
N units of \Pm \\
K D0-branes
\end{center}\end{minipage}}
\put(110,10){\begin{minipage}[t]{4cm} \begin{center}
DLCQ M on  $\mathbf{S}^{1}$\\
N units of \Pm \\
K units of $P'_{10}$
\end{center}\end{minipage}}
\put(48,41){\vector(-1,0){8}}
\put(48,41){\vector(+1,0){8}}
\put(46,44){T$_{1}$}
\put(103,41){\vector(-1,0){8}}
\put(103,41){\vector(+1,0){8}}
\put(101,44){$\beta_{1}$}
\put(48,12){\vector(-1,0){8}}
\put(48,12){\vector(+1,0){8}}
\put(46,15){T$_{1}$}
\put(103,12){\vector(-1,0){8}}
\put(103,12){\vector(+1,0){8}}
\put(101,15){$\beta_{1}$}
\put(18,21){\vector(0,-1){4}}
\put(18,21){\vector(0,+1){4}}
\put(21,20){S} }
\put(73,21){\vector(0,-1){4}}
\put(73,21){\vector(0,+1){4}}
\put(75,20){10-1 flip}
\put(130,21){\vector(0,-1){4}}
\put(130,21){\vector(0,+1){4}}
\put(133,20){S}
\end{picture}
\end{center}
\vspace*{-0.4cm}
\caption{\small The $1+1$ NCOS/DLCQ IIA duality web.
Seiberg's derivation of a non-perturbative Matrix formulation 
for DLCQ IIA (DLCQ M-theory on a transverse $\bS^{1}$) proceeds
along the top (bottom) line and then down to arrive at $1+1$ SYM.
$\beta_{1}$ denotes a boost along $x^{1}$.}
\end{figure}
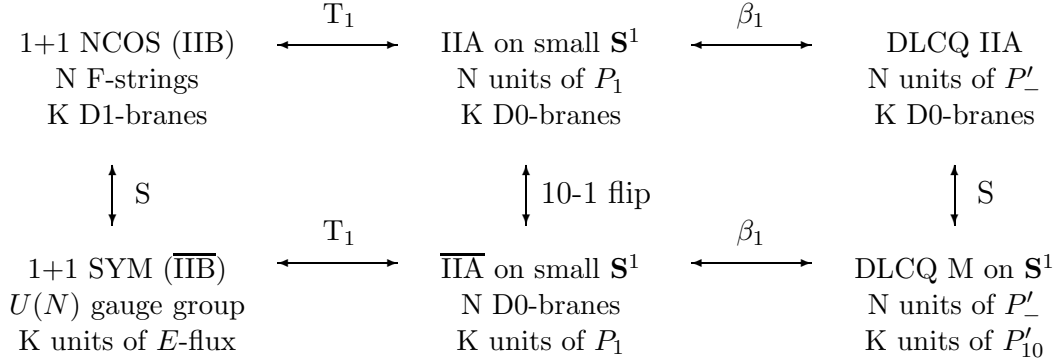

The S-duality between $1+1$ NCOS 
and $1+1$ $U(N)$ SYM with $K$
units of electric flux indicated in Fig.~1 
has been scrutinized in \cite{om,km,hk}.  
As explained in those works, the SYM coupling is
related to the NCOS parameters through\footnote{It is easy 
to see that this is in agreement with the known 
relation to DLCQ IIA parameters \cite{dvv},  
$g^{2}_{YM}=\tRm^{2}/(2\pi\tilde{g}_{s}^{2}\tilde{l}_{s}^{4})$.}
$g^{2}_{YM}=1/(2\pi\ape G_{o}^{4})$.
The effective dimensionless
coupling at energy $E$ is therefore 
$g^{2}_{\mathrm{eff}}\simeq 1/( \ape G_{o}^{4} E^{2})$. At low
energies the theory is strongly coupled,
but the presence of the compact circle
provides an infrared cutoff $1/R$, meaning
that $g^{2}_{\mathrm{eff}}\leq{R^{2}/(\ape G_{o}^{4})}$. 
In Section \ref{ncossec} we will find that this is indeed the
inverse of the effective 
coupling of the NCOS theory, which can be argued to be
$\sqrt{\ape} \Go/R$.

It is interesting to work out the complete
$1+1$ NCOS $\leftrightarrow$ DLCQ IIA
dictionary, which follows from the usual
rules of T-duality. The NCOS open strings are dual to the
usual open strings ending on (and describing the dynamics of)
the IIA D0-brane. NCOS and IIA closed strings are mapped onto one
another, with their winding and Kaluza-Klein numbers interchanged.
(We have already seen one example of this: the $N$ fundamental
strings---i.e., units of electric flux--- appearing in the definition of
the NCOS
theory correspond to the $N$ units of longitudinal IIA momentum,
$\Pm=N/\tRm$.)
The fact that NCOS winding is necessarily positive \cite{km}, $w>0$, is
therefore equivalent to the well-known fact that
in DLCQ longitudinal momentum is
strictly positive, $p'_{-}>0$. The $x^{-}$
winding number in the DLCQ theory is of course arbitrary, just like the
dual NCOS Kaluza-Klein number.
It is worth emphasizing that the standard
interchange of momentum and winding
has in this case the rather non-standard effect of mapping a theory
with gravity to a non-gravitational theory. Of course, in our
post-Matrix/Maldacena era \cite{bfss,malda},
the equivalence of a non-gravitational
theory and a theory of gravity is slightly less surprising.

Klebanov and Maldacena \cite{km} noted that it is possible for
the `NCOS D-string' to emit wound closed strings into the bulk:
in the parent IIB theory, this just corresponds to the $(N,1)$ bound
state dissociating into an $(N-w,1)$ bound state
and a number of fundamental strings with winding numbers $w_{A}>0$
such that $\sum_{A}w_{A}=w$. In the S-dual $\overline{\mbox{IIB}}$
description, one has instead a $(1,N)$ string out of which some
D-strings are separated. In SYM language, this is expressed by
the breaking
$SU(N)\to SU(N-w)\times U(1)^{w}$, keeping the unit of $E$-flux in the
$SU(N-w)$ part.
Now we see that in the T-dual DLCQ IIA
language, this corresponds to the process in which a D0-brane with
$N$ units of longitudinal momentum emits closed strings into the bulk
with $p'_{(A)-}>0$. The cost in light-cone energy that this
process entails
can easily be seen to agree with the one discussed in \cite{km} for
the NCOS theory.
The identification between a $SU(N')$ subsector of
Matrix string theory carrying $K$ units of $E$-flux and $K$
DLCQ IIA D0-branes with total longitudinal momentum $p'=N'/\tRm$
was discussed in \cite{dvv}. 

One of the most surprising aspects of the $1+1$ NCOS theory is the
decoupling of the massless open string modes,
inferred in \cite{km} from the identification of these modes with
the free $U(1)$ sector of the $1+1$ $U(N)$ SYM theory, and verified through
explicit computation of scattering amplitudes with massless open
string vertices. In the Matrix literature it is well-known that
the $U(1)$ part of of the SYM Lagrangian describes the center-of-mass
motion of the DLCQ IIA system in the transverse directions
\cite{bfss,dvv}, 
whose decoupling is obvious. Notice, however, that massless open
string vertices in NCOS are associated only with
quantum fluctuations of the center-of-mass dynamics
of the D1-brane (and the $N$ adsorbed F-strings),
which under T-duality becomes the IIA D0-brane. When
the NCOS or DLCQ IIA state
under consideration includes additional closed strings,
these give rise to a separate contribution
to the center-of-mass modes. Since this contribution is certainly
incorporated in the free $U(1)$ factor of $1+1$ SYM,
there is a question as to what is it that decouples
in NCOS amplitudes which include closed string vertices.
It would be interesting to have a closer look at this point.

{}From the T-dual perspective it is clear that the presence of
the D1-brane in the NCOS theory is not essential: in the DLCQ IIA
theory it is just as
interesting to consider states without any D-branes.
Similarly, it is natural to
ask what DLCQ IIA states with other D-branes correspond
to in the NCOS language. We will elaborate on these issues in Sections
\ref{clstrsec} and \ref{ncossec}.

To complete the dictionary, we should also understand
how the non-commutativity manifests itself
in the T-dual picture. The defining property of NCOS theories is the 
fact that
scattering amplitudes are computed with the usual rules but including
Moyal phases when two operators are interchanged. In the T-dual 
description one
has a D0-brane in motion, which is brought to rest
by means of the (large) boost. 
The boundary conditions for the world-sheet fields
will then be the usual ones 
(Neumann in direction $0$ and Dirichlet in the rest).
The propagator
$\langle X^\mu X^\nu \rangle$ will clearly
not give rise to a phase in this case.

To see what happens consider two open strings with momentum $n^{(1)}/R_1$ and
$n^{(2)}/R_1$ in NCOS prior to taking the limit. These states T-dualize into
open strings ending on the D0-brane with winding number $n^{(1)}$ and
$n^{(2)}$. This calculations are standard (see, e.g., \cite{polchinski}).
Due to the winding number one 
has different left and right momenta,
\be
p_{\mu L,R} = (p_0, \pm\frac{n \tilde{R}_1}{\alpha'},p_\perp)
\ee
 After the boost the D0-brane is at rest but the circle
 is no longer purely spatial. 
The momenta respecting the new periodicity are of the form
\be \label{pLR}
p_{\mu L,R} = (\gamma(p_0\mp\beta\frac{n \tilde{R}_1}{\alpha'}),
 \gamma(\pm\frac{n \tilde{R}_1}{\alpha'}-\beta p_0),p_\perp)
\ee
where $\beta=1-\epsilon/2$ 
is the original speed of the D0-brane and $\gamma=1/\sqrt{1-\beta^2}$.
The propagators have the standard form
\bea
\langle X_\mu(z) X_\nu(w) \rangle &=& -\alpha' \eta_{\mu\nu} \ln(z-w) \\
\langle X_\mu(z) \tilde{X}_\nu(\bar{w}) \rangle &=& \alpha' \delta_{\mu\nu} 
\ln(z-\bar{w})
\eea
where we took into account that $X_0$ obey Neumann boundary 
conditions and the rest Dirichlet.
If one takes a closed string vertex of momentum $p^{(1)}$ around 
another with momentum
$p^{(2)}$ there is a phase change 
$\exp(2\pi i(p^{(1)}_L p^{(2)}_L-p^{(1)}_Rp^{(2)}_R)$,
which for consistency must equal unity. 
However, if one does the same for open string 
vertices, 
as indicated in Fig. 2, there is an extra phase coming 
from the fact that $z$ is also
going (half-way) around $\bar{w}$ 
(and not only around $w$ as in the closed string 
case)\footnote{Here we consider open string vertices to be 
functions of $X(z),\tilde{X}(\bar{z})$ 
with $X(z)$-$\tilde{X}(\bar{z})$ contractions removed. Equivalently,
one can use closed string vertices which are close to the boundary, 
and then the same result follows using OPE's to reduce to open string 
vertices.}.
 This extra phase is given by
\bea
\exp(i \pi \alpha'(p^{(1)}_L p^{(2)}_R - p^{(1)}_R p^{(1)}_L)) &=&
  \exp(4\pi i \beta\gamma^2 \tilde{R}_1(p_0^{(1)} n^{(2)} - p_0^{(2)} n^{(1)}) \\
\eea
Taking the $\epsilon\rightarrow 0$ limit and 
using the T-duality relations one finds that this phase
is finite and given by
\be
 \exp(2\pi i \ape (p_0^{(1)} p_1^{(2)} - p_1^{(2)} p_0^{(2)}))~,
\ee
with $p^{(1,2)}=n^{(1,2)}/R_1$. 
This last expression is the expected Moyal phase, with 
non-commutativity parameter $\theta^{01}=2\pi\ape$.

\begin{figure}[tb]
\centerline{\epsfysize=5cm\epsfbox{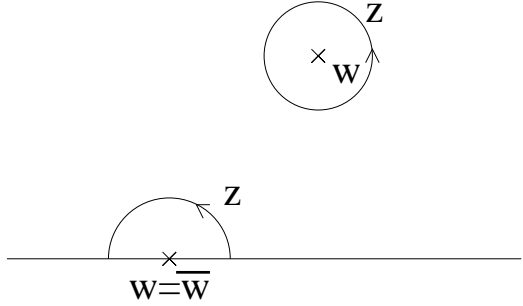}}
\caption{\small A closed string vertex encircling another, and the 
analogous operation for open string vertices, which inverts their order. 
In the latter case there is an extra phase
due to the fact that
$\langle X_\mu(z) \tilde{X}_\nu(\bar{w}) \rangle\neq 0$ .}
\end{figure}

In \cite{km} a general argument
for the vanishing of NCOS amplitudes with at least one massless
open string vertex was presented,
based on the observation that in that case the propagator
is in effect an analytic function, so the 
integral over the position of the vertex
on the boundary can be pulled to the interior and shrunk to zero size. 
{}From (\ref{pLR}) we see that the T-dual statement 
is that for a massless
particle $p_L=0$, which means the vertex is a purely analytic 
function (since $p_L$ is
the momentum multiplying $\tilde{X}(\bar{z})$) and then the 
same reasoning follows. This of
course will be valid in the presence of handles, since the circle can 
be pulled past them; but the argument might fail
if one inserts a closed string vertex operator, for example.

In the Matrix program, DLCQ is not an end in itself:
the ultimate goal is to provide a description
of the corresponding \emph{decompactified} theory.
This is an enormously
difficult task, since it involves understanding
the large $N$ limit of the model in question.
So it is natural to
wonder whether the duality to NCOS theory could afford any new insight
into the nature of this limit. A glance at (\ref{gmmssout2})
is encouraging: $N\to\infty$ is the NCOS weak-coupling limit, $\Go\to
0$ ! Upon closer examination, however, the situation is not as
fortunate: one must remember that in the decompactification limit,
the longitudinal momentum $\Pm=N/\tRm$ (for a state with no
winding) should be held fixed and finite,
and through (\ref{trm}) this translates into the requirement that the
NCOS radius $R$ scale as the inverse of $N$. This is not surprising:
as we have seen, the NCOS description
is related by T-duality to the IIA theory whose decompactified
limit we are trying to understand. The NCOS description is
not particularly transparent in this limit.

In the Matrix context, it has been
argued that the decompactified IIA theory can also be obtained
from a `DLCQ' limit in which $N\propto\eps^{-1}\to\infty$ as $\eps\to
0$, while the radius of
the spatial circle in the (infinite momentum) frame F is held
fixed in string units \cite{bgl,rindep}.
In that case the circle in frame F$'$ has growing radius,
$\tRm\propto\eps^{-1}$, and is not strictly
lightlike. After T-duality (and a change of units),
one arrives at a IIB description with $\ls\propto\sqrt{\eps}$
and \emph{arbitrary} coupling constant \gs.
If $K$ D0-branes are present in the IIA picture, the
$K$ corresponding D1-branes share $N\propto\eps^{-1}$ units of $E$-flux.
The radius $R$ of direction $x^{1}$ is again fixed in string units.
This limit, studied in \cite{verlinde,gkp}, is
distinct from the NCOS limit, although they share some features.
The point we are making is that it must coincide with the large $N$
limit of the NCOS description. That this is surprising is perhaps more
easily appreciated in the S-dual $\overline{\mbox{IIB}}$
description: we are saying that,
as $N\to\infty$, the 
physics of
a system of $N$ D1-branes 
becomes \emph{independent of the string coupling}.
Evidence for this type of independence (in a different context)
was given in \cite{bgl,rindep}. 
This property is to some extent 
responsible for the success of the
Matrix model, and makes it clear
that the large $N$ limit is highly non-trivial.

We should also remark that, whereas Matrix string theory
is conjectured to provide a complete
non-perturbative (second-quantized)
description of DLCQ IIA string theory, the S-dual
$(1+1)$ NCOS picture discussed above utilizes just
the usual perturbative rules of the first-quantized framework.
The same is of course true if one attempts to `directly'
describe the DLCQ IIA theory, along the lines of
\cite{hp,bilal,bilal2,uy}.

\section{Closed Strings, $B$-fields, and Boosts}

\label{clstrsec}

\subsection{The Closed String Spectrum}

\label{clstrspecsec}

As mentioned in the previous section, Klebanov and Maldacena \cite{km}
discovered that upon compactification, the $1+1$ NCOS theory contains
closed
strings with strictly positive winding number. The authors of \cite{km}
determined the spectrum of NCOS closed strings by working in the gauge
where
$B_{01}$ (and not $E$) becomes critical. Let us review their argument.
The
mass formula for closed strings in a $B$-field \cite{polchinski,km} is
\begin{equation}
(p_{0})^{2}+2p_{0}B_{01}{\frac{wR}{\ensuremath{\alpha'}}}-\left(
{\frac{wR}{%
\ensuremath{\alpha'}}}\right) ^{2}(1-B_{01}^{2})-\left(
{\frac{n}{R}}\right)
^{2}-p_{\perp }^{2}-{\frac{2}{\ensuremath{\alpha'}}}(N_{L}+N_{R})=0~,
\label{mass}
\end{equation}
supplemented by the level-matching condition 
\be \label{level}
N_{L}-N_{R}=nw~.
\ee
Interpreting (\ref{mass}) as a mass formula is actually quite confusing: the
nine-dimensional mass depends on the energy $p_{0}$ (and therefore on
$p_{\perp }$). Notice that this is true even of the ten-dimensional mass.
It
is clear that this is a purely stringy phenomenon, given that it only
affects states with non-zero winding, $w\neq 0$. The best way to view
(\ref
{mass}) is as a quadratic equation for the energy. From it one finds the
two
solutions
\begin{equation}
p_{0}=-{\frac{B_{01}wR}{\ensuremath{\alpha'}}}\pm \sqrt{\left(
{\frac{wR}{%
\ensuremath{\alpha'}}}\right) ^{2}+\left( {\frac{n}{R}}\right)
^{2}+p_{\perp
}^{2}+{\frac{2}{\ensuremath{\alpha'}}}(N_{L}+N_{R})}~. \label{p0wB}
\end{equation}
Notice that the quadratic term in $B_{01}$ has canceled inside the
square-root. If we restrict attention to the physical range\footnote{%
This restriction is clear if there are D-branes present: for
$|B_{01}|>1$
there are tachyonic modes in the open string spectrum. In the absence of
D-branes (i.e., in a theory with closed strings only), one can observe
that $%
|B_{01}|>1$ is excluded since  either choice
of sign for the square root would lead to states with
negative energy.  After
T-dualizing it
will be even easier to understand this restriction--- see below.} $%
|B_{01}|\leq 1$, it is clear that we must choose the upper sign in
(\ref{p0wB}) to have $p_{0}>0$.

Let us now examine how (\ref{p0wB}) behaves in the NCOS limit. Writing
$B_{01}=1-\epsilon/2$,
$\ensuremath{\alpha'}=\ensuremath{\alpha'_{e}}\epsilon$
and $p_{\perp}^{2}=k_{\perp}^{2}/\epsilon$, the NCOS limit is
$\epsilon\to 0$
holding \ensuremath{\alpha'_{e}}\ and $k_{\perp}$ fixed. For $w=0$, it
is
easy to see from (\ref{p0wB}) that
$p_{0}\propto\epsilon^{-1/2}\to\infty$,
(unless $k_{\perp}=N_{L}=N_{R}=0$). For $w\neq 0$, the NCOS limit turns
(\ref
{p0wB}) into
\begin{equation} \label{p0ncos}
p_{0}={\frac{R}{\epsilon\ensuremath{\alpha'_{e}}}}(-w+|w|)+{\frac{wR}{2%
\ensuremath{\alpha'_{e}}}} +{\frac{\ensuremath{\alpha'_{e}}}{2|w|R}}%
k_{\perp}^{2} +{\frac{N_{L}+N_{R} }{|w|R}} + {\mathcal{O}}(\epsilon)~.
\end{equation}
Clearly the energy diverges unless we arrange for the first term to drop
out, i.e., if $w>0$. In that case we find
\begin{equation} \label{p0km}
p_{0}={\frac{wR}{2\ensuremath{\alpha'_{e}}}}
+{\frac{\ensuremath{\alpha'_{e}}%
}{2wR}}k_{\perp}^{2} +{\frac{N_{L}+N_{R} }{wR}},
\end{equation}
which indeed agrees with the result of \cite{km} (obtained by noting
that in
the NCOS limit (\ref{mass}) turns into a linear equation--- this is true
only if one assumes that $p_{0}$ is finite in the limit). A whole tower
of
closed string excitations is thus seen to be present in the NCOS theory.
These states show up as poles in NCOS open string scattering amplitudes
\cite
{km}.

\subsection{$B$ is for Boost}

\label{boostsec}

While the approach of Klebanov and Maldacena \cite{km} reviewed in the
previous subsection yields the correct spectrum (i.e., the one seen in
open
string scattering), it is somewhat mysterious that the observed
decoupling
appears to depend on the presence of the $B$-field, or in other words,
on
the choice of a particular gauge. In the gauge where $B=0$ and it is $E$
that becomes critical, the (free) closed string spectrum must be the
standard one, because closed strings are electrically neutral. In this
subsection and the next we will clarify this issue.

We start from the following observation. It is well-known that a
D$p$-brane
with a constant electric field $E$ is equivalent a D$(p-1)$-brane moving
at
speed $v=E$: the two are mapped onto one another under T-duality along
the
direction of the field. It should therefore be possible to transform
away a
constant electric field through a T-duality transformation followed by
an
appropriate boost and a new T-duality. Since we also know that a gauge
transformation can exchange an electric field for a $B_{01}$-field, we
conclude that the effect of the latter field on a closed string can be
understood by boosting a system with $B=0$.

Let us now fill in the details of the above argument. We begin with the
IIB theory with $x^{1}$ compactified on a circle of radius $R$ and with
$B=0$. After a simple T-duality exchanging
the winding and momentum, the energy formula for a closed string 
becomes
\begin{equation}
p_{0}=\sqrt{\left(
{\frac{n\ensuremath{\tilde{R}}}{\ensuremath{\alpha'}}}%
\right) ^{2}+\left( {\frac{w}{\ensuremath{\tilde{R}}}}\right)
^{2}+p_{\perp
}^{2}+{\frac{2}{\ensuremath{\alpha'}}}(N_{L}+N_{R})}~, \label{p0nob}
\end{equation}
where $\tilde{R}=\alpha^{\prime}/R$ is the radius of the circle
in the IIA description, whereas $n$ and $w$ are the original IIB 
Kaluza-Klein and winding numbers. 

If we perform a boost with velocity $v$ along the $x^{1}$ direction, the
energy in the boosted frame F$^{\prime }$ (where the circle is no longer
purely spatial) will read
\begin{equation} \label{p0boost}
p_{0}^{\prime }=-\gamma v{\frac{w}{\ensuremath{\tilde{R}}}}+\gamma
\sqrt{%
\left( {\frac{n\ensuremath{\tilde{R}}}{\alpha ^{\prime }}}\right)
^{2}+\left( {\frac{w}{\ensuremath{\tilde{R}}}}\right) ^{2}+p_{\perp
}^{2}+{%
\frac{2}{\alpha ^{\prime }}}(N_{L}+N_{R})}~, \label{p0'}
\end{equation}
where $\gamma =1/\sqrt{1-v^{2}}$. We now wish to compare this
with the T-dual image of (\ref{p0wB}), which describes
a theory with a $B$-field, compactified on a purely
spatial circle of radius $R$.
For this we apply the standard rules for T-duality in the presence
of a $B$-field--- see, e.g., \cite{gpr}. Defining
\[
E_{ab}=g_{ab}+B_{ab}=\left(
\begin{array}{cc}
-1 & R B_{01} \\
-R B_{01} & R^{2}
\end{array}
\right)
\]
(in coordinates such that direction $1$ has period $2\pi$),
and transforming according to
\[
E\rightarrow \left( aE+b\right) \left( cE+d\right) ^{-1}~,
\]
with
\[
a=d=\left(
\begin{array}{cc}
1 & 0 \\
0 & 0
\end{array}
\right) \mathrm{\ \ \ ,\ \ \ }b=c=\left(
\begin{array}{cc}
0 & 0 \\
0 & 1
\end{array}
\right)~,
\]
leads to a new metric
\be \label{tdualmetric}
g_{ab}^{\prime \prime }=\left(
\begin{array}{cc}
-1+B_{01}^{2} & -B_{01}/R \\
-B_{01}/R & 1/R^{2}
\end{array}
\right)
\ee
and no $B$-field\footnote{For related results see \cite{mp}.}. 
In the case of compact Euclidean time the result of such
a T-duality is to map a straight torus with a $B$-field into a tilted torus
without a $B$-field. The spectrum is of course invariant under this 
transformation, so it is still given by the mass formula
(\ref{mass}), except that $w$ and $n$
are interpreted in the IIA theory
as momentum and winding, respectively. One should
note that the second term, linear in $B_{01}$, is no 
longer due to winding and
the presence of a $B$-field, but to a mixing of momentum 
and energy resulting from the non-diagonal metric (\ref{tdualmetric}). 
{}From (\ref{mass}) we deduce
that the energy is given by
\begin{equation}
p_{0}^{\prime \prime }=-B_{01}{\frac{w}{\ensuremath{\tilde{R}}}}+\sqrt{%
\left( {\frac{n\ensuremath{\tilde{R}}}{\alpha ^{\prime }}}\right)
^{2}+\left( {\frac{w}{\ensuremath{\tilde{R}}}}\right) ^{2}+p_{\perp
}^{2}+{%
\frac{2}{\alpha ^{\prime }}}(N_{L}+N_{R})}~.
\end{equation}

The final step is to compare this with (\ref{p0'}). The frame F$''$
in which the off-diagonal metric is written is such that one coordinate
axis is along the compact direction, just as before the T-duality, even
though this direction is no longer purely spatial. 
The other axis is along the original time direction. 
To compare with (\ref{p0'}) we must change
coordinates to
those of the boosted coordinate system F$'$, bringing the metric back to a
diagonal
form. Applying the necessary transformation to the energy and momentum
in frame F$''$
one finds
\[ \left(\begin{array}{c}
p'_{0} \\
 p'_{1}
\end{array}\right)=
{1\over\sqrt{1-B_{01}^{2}}} \left(
\begin{array}{cc}
1 & 0 \\
-B_{01} & 1-B_{01}^{2}
\end{array}
\right) \left(
\begin{array}{c}
p_{0}-B_{01}p_{1} \\
p_{1}
\end{array}
\right) =\left(
\begin{array}{c}
\gamma p_{0}-\gamma vp_{1} \\
-v\gamma p_{0}+\gamma p_{1}
\end{array}
\right)
\]
where in the second step
we have identified $v=B_{01}$. This precisely adds the factor of
$\gamma $ present in (\ref{p0'}), and so completes the analysis.
The identification of $B_{01}$ and $v$ is the natural analogue of the
fact
that under T-duality one has $E=2\pi \ensuremath{\alpha'}F_{01}%
\leftrightarrow v=\partial _{0}X^{1}$ for a D-brane. In particular, it
leads
us to conclude that $B_{01}$ has maximum value one--- even in the
absence of
D-branes.

Returning to the NCOS setting, an immediate observation 
is that
 the description where $B$ (and not $E$%
) is made critical is the one that is more `directly' T-dual to DLCQ
IIA.
Remember that this is the one description where one can easily work out
the
spectrum of closed strings, as in \cite{km}. Open strings are of course
only
sensitive to the combination $E+B$.

\subsection{The Decoupling}

\label{decouplingsec}

We have just learned that the gauge-transformation which shifts $F_{01}$
into $B_{01}$ is T-dual to a boost. It is clear then that, when the
$x^{1}$
direction is compact, this (large) gauge transformation, just like the
boost, \emph{does not} leave the spectrum invariant. On the other hand,
it
is evident that the physics must be the same in either gauge, since
these
are simply different descriptions of a single physical system. Now, as
we
have reviewed, in the $(E=0,B\neq 0)$ gauge the decoupling of closed
strings
with non-positive winding number can be easily seen to follow from the
energetics \cite{km}. In the $(E\neq 0,B=0)$ gauge, on the other hand,
there
is no dependence of the closed string (free) spectrum on $E$, because
closed
strings are electrically neutral. We thus appear to have a
contradiction:
how could a decoupling take place in this gauge?

Let us examine the situation more closely. The standard mass-shell
condition
for a closed string (in the absence of $B$) implies that
\begin{equation}
p_{0}=\sqrt{\left( {\frac{wR}{\ensuremath{\alpha'}}}\right) ^{2}+\left(
{%
\frac{n}{R}}\right) ^{2}+{\frac{k_{\perp }^{2}}{h}}+{\frac{2}{%
\ensuremath{\alpha'}}}(N_{L}+N_{R})}~. \label{p0}
\end{equation}
We would like to extract from this expression a decoupling of all $w\leq
0$
states in the NCOS limit: $\epsilon \rightarrow 0$ with
$\ensuremath{\alpha'}%
=\epsilon \ensuremath{\alpha'_{e}}$, $h=\epsilon $
($\ensuremath{\alpha'_{e}}%
,k_{\perp }$ fixed). It is obvious that the scaling of the electric
field on
the D1-brane, $E=1-\epsilon /2$, is irrelevant. We see from (\ref{p0})
that
in the $(E\neq 0,B=0)$ gauge we would expect \emph{all} closed string
states
to decouple! In particular, at this level there does not seem to be any
difference between states with positive and negative winding.

Inferring decoupling in this manner is clearly too naive. 
Eq.~(\ref{p0}) certainly implies that, in the NCOS limit, the energy of
closed
strings with any winding number is much higher than that of open strings
(the latter is of order $1/\sqrt{\ape}$). We should remember,
however,
that the open strings describe excitations of an $(N,1)$ string, whose
rest
energy also should be taken into account. It is given by the standard
BPS
formula,
\begin{equation}
2\pi RT_{(N,1)}={\frac{R}{\ensuremath{\alpha'}}}\sqrt{N^{2}+{\frac{1}{%
\ensuremath{g_s}^{2}}}}~, \label{n1string}
\end{equation}
which clearly diverges in the NCOS limit,
\begin{equation}
2\pi RT_{(N,1)}\simeq {\frac{R|N|}{\ensuremath{\alpha'_{e}}\epsilon }}%
\rightarrow \infty ~. \label{n1stringncos}
\end{equation}
Starting with the $(N,1)$ string (excited or not), there is thus plenty
of
energy available to create closed strings.

To make a fair comparison, then, we should allow for the appropriate
subtraction also in the case of the closed strings. In this context it
is
important to remember that the sum of the number of electric flux units
on
the D1-brane and the closed string winding number, $W=N+w$, is conserved
by
interactions. When we compare energies of states with different values
of $w$%
, we should consequently make sure that they belong to the same $W$
sector--- otherwise the comparison is meaningless. Let us then fix a
(positive) value of $W$ once and for all. {}From (\ref{p0}) and (\ref
{n1string}), it follows that (ignoring interactions) a state with the
$(N,1)$
string and a closed string with $w=0$ has total energy
\begin{equation}
P_{0}=\epsilon ^{-1}{\frac{R|W|}{\ensuremath{\alpha'_{e}}}}+{\epsilon
^{-1/2}%
}\sqrt{k_{\perp
}^{2}+{\frac{2}{\ensuremath{\alpha'_{e}}}}(N_{L}+N_{R})}+{%
\mathcal{O}}(1)~, \label{P0now}
\end{equation}
whereas if the closed string is wound, $w\neq 0$, the total energy of
the
state is
\begin{equation}
P_{0}=\epsilon
^{-1}{\frac{R(|W-w|+|w|)}{\ensuremath{\alpha'_{e}}}}+{\frac{%
|W-w|R}{2\ensuremath{\alpha'_{e}}}}+{\frac{\ensuremath{\alpha'_{e}}k_{\perp
}^{2}}{2|w|R}}+{\frac{N_{L}+N_{R}}{|w|R}}+{\mathcal{O}}(\epsilon )~.
\label{P0w}
\end{equation}
It should be noted that in the second term of (\ref{P0w}) we have made
use
of the relation $\ensuremath{g_s}(W-w)=\sqrt{\epsilon }$, which follows
from
the quantization condition (\ref{quant}) for the near-critical electric
field on the brane, $E=1-\epsilon /2$.

The first term in (\ref{P0now}) is the `rest energy' due to $W$, so it
is
evidently common to all states in the sector under
consideration. Subtracting it, we see from (\ref{P0now}) and (\ref{P0w})
that the `dynamical' energy diverges like $\epsilon^{-1/2}$ if $w=0$,
diverges like $\epsilon^{-1}$ if $w<0$, and is finite only if $w>0$ (and
$w<W$). The decoupling of all $w\leq 0$ states is now evident. In the $w>0$
case, the dynamical energy is given by
\begin{equation} \label{P0wpos}
P_{0}= {\frac{(W-w)R}{2\ensuremath{\alpha'_{e}}}} +{\frac{%
\ensuremath{\alpha'_{e}} k_{\perp}^{2} }{2wR}}+{\frac{N_{L}+N_{R}
}{wR}}~,
\end{equation}
which agrees with the result (\ref{p0km}) of Klebanov and Maldacena
\cite{km}, 
except that $w$ has been replaced with $N=W-w$ in the numerator of the
first term. This difference is partly due to the inclusion of the
`dynamical' energy of the $(N,1)$ string.

We have thus succeeded in understanding the decoupling of closed strings
with non-positive winding directly in the $(E\neq 0,B=0)$ gauge.
Before closing this section, let us address another possible point of
concern: given that the NCOS limit takes $\ensuremath{g_s}\to\infty$, do
we
really learn anything from (\ref{p0}) or (\ref{p0wB}), which describe
the
closed string spectrum of the \emph{free} IIB theory? Notice that this
is
not an issue of continuing to strong coupling, since we are considering
the
full spectrum, not just BPS states. The essential point is that the
\emph{effective} closed string coupling constant for
closed strings is not $\ensuremath{g_s}$, but 
$\Go\sqrt{\ape}$,
which remains finite in the limit.
This
has been shown in \cite{ncos} by determining the cost of adding a handle
to
a worldsheet describing open string scattering to be
$G_{o}^{4}\ape$.
In the next section we will understand this result from a different
perspective.

\section{IIA/B Wound String Theory} \label{ncossec}

\subsection{Closed Strings} \label{ncosclstrsec}

In the previous section we have seen that the decoupling of closed
strings with non-positive winding number in the $1+1$ NCOS theory can
be made explicit even in the absence of a $B$-field. The effect clearly
does not depend on the presence of the D1-brane, either.
So we learn that, as long as there is a compact dimension,
one can meaningfully define an `NCOS' limit of IIB (or IIA) string theory
even in a vacuum with zero background fields and no D-branes.
To be more precise, we consider IIA/B string theory compactified on a 
circle of radius $R$ along direction $x^{1}$, with metric 
(\ref{metric}), in the limit
\be \label{woundlim}
\delta\to 0, \quad
\gs={\Gs\over\sqrt{\delta}},\quad  
\ls=\Ls\sqrt{\delta}, \quad
h=\delta, \quad 
\mbox{with}\quad \Gs,\Ls,R\quad \mbox{fixed}.
\ee
We have defined here a rescaled coupling constant and string 
length; their relation to the parameters $\Go,\ape$ employed in 
\cite{sst2,ncos,om} and in the 
previous sections will be discussed in Section \ref{dbranesec}.
We now proceed to examine the theory obtained in the limit 
(\ref{woundlim}), which contains only closed strings.

To begin with, it is worth revisiting the decoupling argument
from a slightly different perspective.
Consider a scattering process in which
the initial state has $C$ closed strings with winding numbers
$w_{I}$ and momenta $p_{\mu I}=(p_{0I},n_{I}/R,p_{\perp I})$,
where $I=1,\ldots,C$ and the
coordinates are such that the metric is given by (\ref{metric}).
The corresponding quantities in the final state will be denoted by
primed variables.
As we have seen before, in the limit (\ref{woundlim})
with $p_{\perp}$ fixed, the energy of the
individual strings diverges: it is
\be \label{p0unwound}
p_{0}= {\delta^{-1/2}}
\sqrt{p_{\perp}^{2}+{2\over\Ls^{2}}(N_{L}+N_{R})}
+\cO(1)
\ee
for an unwound closed string, and
\be \label{p0wound}
p_{0}= \delta^{-1}{|w|R \over \Ls^{2}}
+{\Ls^{2} p_{\perp}^{2} \over 2|w|R}+{N_{L}+N_{R} \over |w|R}
+\cO(\delta)
\ee
for a string with $w\neq 0$.

The important observation is that,
for energy to be conserved in the scattering process, it must
do so separately at each order in an expansion in powers
of $\delta$. Equating the coefficient
of $\delta^{-1}$ in the initial and final energies we obtain the
requirement
\be \label{encons}
|w_{1}|+\ldots+|w_{C}|=|w'_{1}|+\ldots+|w'_{C'}|~.
\ee
On the other hand, the statement of
winding number conservation reads
\be \label{wcons}
w_{1}+\ldots+w_{C}=w'_{1}+\ldots+w'_{C'}~.
\ee
Now, suppose all of the initial strings have positive winding,
$w_{I}>0$. Then the only
way that (\ref{encons}) and (\ref{wcons}) can be simultaneously
satisfied is if all of the final strings also have non-negative
winding, $w'_{I}\ge 0$. Moreover, (\ref{p0unwound}) shows that
if $w'_{I}=0$ for some $I$ then there would be a contribution of
$\cO(\delta^{-1/2})$ to the final energy, which would have no
counterpart in the initial energy.
We therefore conclude that
no strings with non-positive winding can
be produced in the scattering event.

Notice that the restriction on the \emph{initial} state was put
in by hand. Clearly, if some of the incoming strings
have negative (or zero) winding, then these \emph{can}
interact with positively-wound
strings. The important point is that one can consistently
restrict attention to the subsector where all closed strings have 
strictly positive winding\footnote{To be precise, we should note that
unwound closed strings can be
considered as long as they have $N_{L}=N_{R}=p_{\perp}=0$. Their
energy is then finite in the limit (\ref{woundlim}), 
$p_{0}=|n|/R$. Such strings
would clearly not make any non-trivial contribution to the dynamics,
so they can be ignored.}, 
and within it,
conservation of the leading, $\cO(\delta^{-1})$ piece
of the energy is guaranteed by winding-number conservation.
It is thus sensible to define a `dynamical' energy
by removing this divergent contribution:
$\hat{p}_{0}=p_{0}-wR/\delta\ap$.
The total dynamical energy is conserved and finite.

Given the fact
that they contain only (positively) wound strings,
it is natural to call these theories the Type II
Wound String Theories (or Wound IIA/B, for short). We will consequently
refer to the limit (\ref{woundlim}) as the Wound limit. 

It should be noted that
the decoupling argument we have given above mirrors the standard
discussion in the infinite-momentum frame (IMF)
\cite{weinberg,ks}. 
This is no accident: as we have seen in
Section \ref{dlcqsec},
(ordinary) T$_{1}$-duality
maps the Wound IIA/B theory onto IIB/A
in the IMF (with a compact longitudinal
direction\footnote{We are employing the term IMF here
to refer to the frame F where the circle is purely
spatial: the momentum is infinite in this frame
as a result of the shrinking radius.
This frame is related to the DLCQ frame F$'$ by a large boost.}),
with the Wound winding $W$ determining the IMF longitudinal momentum,
$P_{1}=W/\tR\to\infty$.
This is obviously conserved, so it is natural
again to subtract its (divergent) contribution to the energy. The
remaining `dynamical' energy is precisely what
corresponds to the finite \emph{lightcone} energy $\Pp$ in the
the frame F$'$ where the circle is
null (and $\Pm=W/\tRm$ is finite).
We also understand by now that the role of the near-critical
$B_{01}$-field employed in \cite{km} is merely to effect this
subtraction.
The restriction to the lowest energy subsector of the IMF theory
is standard;
it is necessary to match the
degrees of freedom of a direct DLCQ description (where all momenta
are strictly positive from the beginning) \cite{my,pb}.

The dynamical energy of a Wound string is
\be \label{p0dyn}
\hat{p}_{0}={\Ls^2 p_{\perp}^{2} \over 2wR}+{N_{L}+N_{R} \over
wR}~.
\ee
Notice that, compared to the Klebanov-Maldacena result, Eq.~(\ref{p0km}),
the above expression is missing the first term. That term
represents a rescaled winding energy, and it can be seen to play the role
of a Newtonian mass in the non-relativistic-type
expression (\ref{p0dyn}). But again,
since the total winding number (just like the total Newtonian mass)
is conserved, inclusion of this term, albeit natural,
does not affect the dynamics: it
constitutes an irrelevant shift of the energy.
The naive string tension inferred from this Newtonian mass is 
\begin{equation}  \label{ncosF1}
\hat{T}_{F}= \frac{1}{2\pi\Ls^{2}}~.
\end{equation}
This agrees with the usual result, and consequently differs from the
tension discussed in \cite{km} by a factor of two. In other words, 
(\ref{p0km}) is a little peculiar: the first term is only 
half of the Newtonian mass inferred from the second term. 
Even so, it was shown in \cite{km} that
(\ref{p0km}) is in exact agreement with the corresponding $1+1$ SYM 
result. A further 
peculiarity of (\ref{p0dyn}) is the presence of the second
term, which amounts to a contribution to the rest mass of the string
distinct from the Newtonian mass discussed above.

To summarize,
we have shown that upon compactification on a (purely spatial)
circle, there exists a meaningful `NCOS' limit
of IIA/B string theory in the absence of D-branes. This yields a
theory of closed strings with strictly positive winding, Wound IIA/B,
which is T-dual to DLCQ IIB/A.
Agreement between the corresponding scattering amplitudes is
guaranteed by the usual rules of T-duality--- we will elaborate on
this point in the next subsection.

\subsection{Closed String Amplitudes} \label{amplsec}

Since the closed string sector of Wound IIA/B is T-dual to
that of DLCQ IIB/A, we expect the scattering amplitudes
of these closed strings to be well-defined in the Wound
limit (\ref{woundlim}).
The energy formula (\ref{p0dyn}) plus conservation of winding number
shows that the system can be described as non-relativistic particles
with masses\footnote{These particles can be thought of as bound 
states of $w_{I}$ particles with winding number one.} 
$w_I R$ and an internal energy with the spectrum of a
harmonic oscillator. In addition, there is an extra conserved charge $n$,
the momentum along the circle, which appears only through the constraint
(\ref{level}).
These particles move in eight transverse dimensions and their
scattering is given by the Wound limit of the usual scattering
amplitudes for strings wound around a circle 
(see, e.g., \cite{polchinski}).

When string theory is compactified on a circle one has to distinguish
between left- and right-moving
momenta along the compact direction, defined as
\be
k_{L,R} = \frac{n}{R} \pm {w R\over\ap}~.
\ee
A generic amplitude depends on the
Lorentz invariant products $\alpha' k_{iL,R} k_{jL,R}$,
and the important observation is that
these are finite in the limit. For example, in the case
of external particles with zero momentum along the circle
one has $k_{iL} k_{jL}=k_{iR} k_{jR}$, which in the
limit reduces to
\bea
\alpha' k_{iL} k_{jL} &=& \frac{2w_iw_jR^2}{\alpha'} +
\frac{w_i}{w_j}\left(\Ls^2(k_j^\perp)^2+\mu_j^2\right)
+ \frac{w_j}{w_i}\left(\Ls^2(k_i^\perp)^2+\mu_i^2\right)  
\nonumber \\
&& - 2\Ls^2 k_1^\perp k_j^\perp - \frac{2 w_iw_jR^2}{\alpha'}
\nonumber \\
&=& \mu_j^2\frac{w_i}{w_j} +
\mu_i^2\frac{w_j}{w_i}
+ \Ls^2 w_iw_j\left(
\frac{1}{w_j}k_j^\perp-\frac{1}{w_i}k_i^\perp\right)^2~,
\eea
where $\mu^{2}_{j}=2(N_{L}+N_{R})$,
and we have taken into account the fact that the transverse metric is
$g_{ij}=(\Ls/\ls)^{2} \delta_{ij}$.
Note that the dangerous terms that behave as $1/\alpha'$ cancel.
This means that one can safely take the Wound limit (\ref{woundlim}). 
For example, in the case of a
four-point amplitude of massless particles, the Mandelstam variables are
\bea \label{eq:stu}
\alpha' s &=& -(k_{1,L} + k_{2,L})^2
= \Ls^2 w_1w_2\left(
\frac{1}{w_2}k_2^\perp-\frac{1}{w_1}k_1^\perp\right)^2
= \Ls^2 w_1w_2 (v_1^\perp-v_2^\perp)^2
 \\
\alpha' t &=& -(k_{1,L} - k_{3,L})^2
= \Ls^2 w_1w_3\left(
\frac{1}{w_3}k_3^\perp-\frac{1}{w_1}k_1^\perp\right)^2
= \Ls^2 w_1w_2 (v_1^\perp-v_3^\perp)^2
\nonumber \\
\alpha' u &=& -(k_{1,L} - k_{4,L})^2
= \Ls^2 w_1w_4\left(
\frac{1}{w_4}k_2^\perp-\frac{1}{w_1}k_1^\perp\right)^2
= \Ls^2 w_1w_2 (v_1^\perp-v_4^\perp)^2 \nonumber 
\eea
where $v^\perp_i=k_i^\perp/w_i$ is the transverse velocity of the
string.
In the Wound limit
the form of the amplitude is left intact; the only modification
is to replace $s,t,u$ by the formulas (\ref{eq:stu}). For instance, one can
take two strings with winding number $w_1$, $w_2$, a relative
velocity $v_{12}$ in direction $\hat{8}$, and relative momentum $p_{12}$
in the other $7$ dimensions. Expanding for small $|p_{12}|$ and
Fourier transforming to position space one would obtain the dimensional 
reduction of the usual
$v_{12}^4/r^7$ potential for two D0-branes, in a manner similar
to \cite{harvey}.
Here we do not need to take $v_{12}$ small, since that is included as
part of the Wound limit.

Let us now examine what happens with one-loop corrections.
As shown in \cite{bilal}, in the T-dual language the torus amplitude
has a finite limit. In our example of four massless particles, this
one-loop amplitude is given by \cite{bilal,rindep}
\bea
A^{(4)} &=& K \int \frac{d^2\,\tau}{(\Ima \tau)^2}\int\prod_{a=1}^{3}
\frac{d^2\,\nu_a}{\Ima\tau} {\cal I} \\
{\cal I} &=& \prod_{a,b} E(\nu_{ab},\tau)^{\alpha'k_{aR}.k_{bR}/2}
\bar{E}(\nu_{ab},\tau)^{\alpha'k_{aL}.k_{bL}/2}
e^{\frac{\pi\alpha'}{\Ima\tau}\left(\sum_ak_a^\perp\Ima\nu_a\right)}
{\cal S} \nonumber \\
{\cal S} &=& g_s^2 \sqrt{\Ima \tau} \frac{\sqrt{\alpha'}}{R}
\sum_{n,m}
e^{-\pi\Ima\tau\left(\frac{m^2R^2}{\alpha'}+\frac{\alpha'n^2}{R^2}\right)
-2\pi imn\Rea\tau} \times  \nonumber \\
&&\times e^{-i\pi\left(\frac{\alpha'}{R}n(k^1_{La}\bar{\nu}_a-k^1_{Ra}\nu_a)
+mR(k^1_{La}\bar{\nu}_a+k^1_{Ra}\nu_a)\right))} \nonumber 
\eea
where $E(\nu_{ab},\tau) = \Theta_1(\nu|\tau)/\Theta'_1(0|\tau)$ and
$\nu_{ab}=\nu_a-\nu_b$. Also we included in $K$ the kinematic factor
and all the constants which are not important for our purpose.

The function ${\cal S}$ is manifestly T-duality invariant, including the
coefficient $g^2_s \sqrt{\alpha'}/R$ in front.
In order to take $\alpha'\rightarrow 0$ it is convenient to
perform a Poisson resummation on $n$, which yields \cite{bilal}
\be
{\cal S} = g_s^2 \sum_{m,n}
e^{-\frac{\pi R^2}{\alpha'\Ima\tau}
\left\{(n+m\tau-\frac{\alpha'}{R}k^1_{Ra}\nu_a)
(n+m\bar{\tau}+\frac{\alpha'}{R}k^1_{La}\bar{\nu}_a)
+\frac{\alpha'^2}{4R^2}(k^1_{La}\bar{\nu}_a+k^1_{Ra}\nu_a)^2\right\}}~.
\ee
Using $k^1_{L,Ra} = \pm m_{a}R/\alpha'$ we obtain
\be
{\cal S} = g_s^2 \sum_{m,n} e^{-\frac{\pi R^2}{\alpha'\Ima\tau}
|n+m\tau-m_a\nu_a|^2}~.
\ee
As argued by Bilal \cite{bilal}, in the $\alpha'\rightarrow 0$ limit
this becomes a delta-function, 
\be \label{torusamp}
{\cal S} = g_s^2 \alpha' \frac{\Ima\tau}{R^2}\sum_{m,n}\delta^{(2)}
(n+m\tau-m_a\nu_a)~,
\ee
which converts one of the $\nu_a$ integrals into a sum.
Then, when $\alpha'\rightarrow 0$ the amplitude vanishes unless we
scale $g_s \sim \Gs/\ls\to\infty$,
as in the Wound limit (\ref{woundlim}).
We have thus shown
that the argument in \cite{ncos} stating that the handles
are weighted by $G_s^{2}$ and not by $g_s^2$ has a T-dual counterpart in
the argument of \cite{bilal}.

\subsection{D-branes and Couplings} \label{dbranesec}

Having understood the effect of the
Wound limit (\ref{woundlim})
on closed strings from various different perspectives,
let us now re-examine its effect on D-branes.
As discussed in Section \ref{decouplingsec},
the energy of a D1-brane wrapped on the Wound circle which carries
$N$ units of electric flux is given by (\ref{n1string}). For $N\neq 0$
this becomes
\be \label{N1}
2\pi R T_{(N,1)}= {R |N| \over \Ls^2\delta}+
{R\over 2|N|\Gs^{2}\Ls^2}+\cO(\delta)
\ee
in the Wound limit. 
The first term is of the same order as
the divergent term in (\ref{p0wound}), reflecting the fact
that electric flux on the brane is equivalent to fundamental string
winding. In the remainder of the paper,
we will find it most convenient to use the latter terminology.
Again, the key point is that as long as $N>0$, one can
subtract the divergent term on account of conservation of the
(total) winding number. We thus learn that D-strings exist in the
Wound theory only if they have strictly positive (F-string)
winding.
This again is easy to understand in the T-dual language: only
D0-branes with strictly positive longitudinal momentum are part of
the DLCQ IIA theory.  

It is by now evident that the various setups
which have hitherto been known as `$(p+1)$-dimensional
NCOS theories'  
\cite{sst2,ncos,om}
are simply different truncations of the state space of the corresponding
Wound theory to the subspace of states containing a D$p$-brane
extended along the Wound direction. 
We also understand that the need to turn on 
an electric field in the D$p$-brane
worldvolume (which then becomes critical in the limit) is merely an 
expression of the fact that only objects carrying strictly positive 
F-string winding remain in the spectrum\footnote{We should note that 
this requirement also applies to NS5-branes, which will be present in 
the spectrum of the theory only if they carry positive fundamental 
string winding.}. 

We are now ready to address the question of how the Wound closed string
parameters $\Gs,\Ls$ introduced in (\ref{woundlim}) are related to the
NCOS open string quantities employed in \cite{sst2,ncos,om} 
and Sections \ref{dlcqsec} and \ref{clstrsec} of the present paper.
In the standard NCOS discussion one infers from the formulas
of \cite{sw,gkp} and the electric-flux quantization condition (\ref{quant})
that open strings describing the dynamics of a single
D1-brane carrying $N$ units of F-string winding 
interact with strength $G_{o}^{2}=1/N$, and their tension is set 
by a parameter $\ape$.  With this as a
starting point it might be tempting to define a closed string coupling
using the standard formula $G_{s}=G_{o}^{2}$. 
However, in our context this would clearly not be an appropriate
definition, because $\Go$ is a function of
the parameters $N$ and $p$ that characterize a specific brane 
configuration of the
theory, whereas the closed string coupling
$\Gs$ should be universal. Related to this, the rescaling of the 
transverse metric (\ref{metric})
in the standard NCOS limit (\ref{gmmsslim}) is in fact also $p$- and 
$N$-dependent. For instance, for $p=1$ the quantization condition 
(\ref{quant}) implies that $\eps=1/(N\gs)^{2}$, as seen in (\ref{gmmssout}).
Again, whereas this is not unreasonable for the purpose of 
\cite{sst2,ncos,om}, which is 
to consider one brane configuration at a time, it is not convenient 
now that we have come to realize that these are all just different 
states in the same underlying theory.

The essential content of the limits (\ref{gmmsslim}) and 
(\ref{woundlim}) is the scaling $\gs\to\infty$, $\ls\to 0$, holding
$\gs\ls$ fixed. {}From this observation it follows that 
\be \label{ncostowound}
\Gs\Ls=\gs\ls=\Go\sqrt{\ape}~.
\ee
It is important to understand
that in the right-hand side the dependence of the parameters
$\Go$ and $\sqrt{\ape}$ on the specific brane 
quantities $N$ and $p$ cancels out. Similarly, taking into account 
the different rescalings of the metric we infer that the perpendicular 
momenta in the two descriptions are related through
\be \label{pperp}
\Ls^{2}p_{\perp}^{2}=\ape k_{\perp}^{2}~.
\ee
We have in fact already made implicit
use of this relation when comparing 
(\ref{p0dyn}) with (\ref{p0km}).

In a certain sense, 
the parameter that truly characterizes the closed string sector is the 
\textit{dimensionful} quantity (\ref{ncostowound}),
which is held fixed in the Wound limit. 
For a perturbative expansion this dimensionful coupling should be compared
with some other relevant scale in order to form a dimensionless
expansion parameter. Since we always have the compactification radius $R$ at
our disposal, it makes physical
sense to focus on $\gs\ls/R$. That this is the combination which 
controls the closed string loop expansion from a nine-dimensional
perspective is made evident by the 
result (\ref{torusamp}) for the torus amplitude, which shows the 
addition of a handle costs a factor of $(\gs\ls/R)^{2}$. 
This is not surprising, for $\gs\ls/R$ is in fact the coupling 
constant of the dual DLCQ theory.
It would perhaps seem natural then
to give this quantity the name $G_{s}$, which in view of 
(\ref{ncostowound}) amounts to fixing $\Ls\equiv R$. We choose not to do 
this, however, because it might be a source of confusion, given that
there are processes in the theory where this 
is not the relevant quantity. Examples include adding a handle
to a worldsheet describing
open string scattering, as in \cite{ncos}, or high-energy ($E\gg 1/R$)
closed string scattering, where the relevant expansion parameter is in effect
$\gs\ls E$ (meaning that
the theory is strongly-coupled when $\gs\ls E\gg 1$). 
The fact that the Wound theory lacks
a definite dimensionless coupling is presumably
related to the fact that the Wound limit removes the
dilaton from the spectrum of the theory.

Let us now return to the discussion of the D-branes present in the Wound 
spectrum. It follows from (\ref{N1}) that 
the `dynamical' tension of a Wound D1-brane
with $N$ units of winding is
\be \label{woundD1}
\hat{T}_{D1,N}= { 1\over 4\pi N \Gs^2\Ls^2}~.
\ee
Using (\ref{ncostowound}) and (\ref{gmmssout2}) we can rewrite this as
\be \label{ncosD1}
\hat{T}_{D1,N}= { 1\over 4\pi \Go\ape}~,
\ee
which agrees with the standard formula except for a factor of two 
discussed already in \cite{km}. The form 
(\ref{woundD1}), on the other hand,
has the advantage of manifestly exhibiting the $N$-dependence of the 
tension. Notice that the  heaviest
D-string is the one with a single unit of F-string 
winding. The dependence of (\ref{woundD1}) on $N$, $\Gs$ and $\Ls$ might 
seem peculiar, but it is precisely as needed 
for the energy of the D-string to agree with that of its 
T-dual image in DLCQ IIA. Indeed, using the T-duality formulas 
obtained in Section \ref{dlcqsec}, one finds that
\be
2\pi R \hat{T}_{D1,N}= {\tRm\over 2N} 
\left({1\over\tilde{\gs}\tilde{\ls}}\right)^{2}=\Pp,
\ee
which is the lightcone energy of a D0-brane carrying $N$ units of 
longitudinal momentum.

The preceding discussion can clearly be generalized to
D$p$-branes with $p>1$ which wrap the Wound circle. The simplest way
to do this is to start with the D1-brane carrying $N>0$ units of
(F-string) winding in Wound IIB with coupling constant
$\tilde{\Gs}$, and compactify directions $x^{2},\ldots,x^{p}$ on a
rectangular torus with radii\footnote{These
are held fixed in the Wound limit, so in the closed string metric
relevant for the underlying IIB description the
corresponding proper radii shrink to zero size (they are fixed
in units of $\ls$).} $\tilde{r}_{2},\ldots,\tilde{r}_{p}$.
Ordinary T-duality along these transverse directions \cite{om,kt}
then converts this
into a D$p$-brane wrapped on a torus of radii $R$ and
$r_{i}=\Ls^{2}/\tilde{r}_{i}$, $i=2,\ldots,p$.
Equating the energy of the initial and final
configurations yields the condition
\be \label{tension}
{2\pi R N \over  4\pi N \Gs^2\Ls^2}=
(2\pi)^{p} R r_{2}\cdot\cdot\cdot r_{p}\hat{T}_{Dp}~,
\ee
where $\hat{T}_{Dp}$ denotes the tension of the $p$-brane
after subtraction of the divergent $\cO(\delta^{-1})$ part.
Using the fact that the coupling constant
of the theory after T-duality is \cite{om,kt}
\be \label{tGs}
\Gs=\tilde{\Gs}
\frac{r_{2}\cdot\cdot\cdot r_{p}}{\Ls^{p-1}}~,
\ee
the tension of the brane follows as
\be \label{woundDp}
\hat{T}_{Dp,\nu}= {1\over 2(2\pi)^{p}\nu\Gs^{2}\Ls^{p+1}}~,
\ee
where in the last step we have 
defined a dimensionless parameter
\be \label{nu}
\nu\equiv {N\Ls^{p-1}\over r_{2}\cdot\cdot\cdot r_{p}}~, 
\ee
which expresses the (strictly positive)
\emph{density} of F-string winding carried by the 
$p$-brane. This is clearly the quantity which should be held fixed if 
we wish to take the decompactification limit $r_{i}\to\infty$. 
Again, the tension (\ref{woundDp}) can be written in the more 
familiar but less transparent form
\be \label{ncosDp}
\hat{T}_{Dp,\nu}= {1\over 2(2\pi)^{p} G_{o,p}^{2}\alpha_{e}^{\prime(p+1)/2}}~,
\ee
which uses the effective coupling $G_{o,p}^{2}$ that governs 
interactions of open strings on the $p$-brane, and is related to 
$G_{o}^{2}\equiv G_{o,1}^{2}$ through a formula analogous to 
(\ref{tGs}).

{}It follows from the preceding discussion
that a Wound D$p$-brane extended along $x^{1}$,
with $N$ units of longitudinal (F-string)
winding, is T$_{1}$-dual to a DLCQ IIA/B D$(p-1)$-brane with $N$ units of
longitudinal momentum. One way to see this is to apply the
T$_{2\ldots p}$-duality argument to the known $p=1$ case.
Notice that this procedure
yields only branes which are not wrapped on the null circle.
The DLCQ IIA/B theories also include longitudinally wrapped D$p$-branes,
and these are clearly T$_{1}$-dual to D$(p-1)$-branes in the Wound theory
which are \emph{transverse} to the Wound circle.
The simplest case is that of
the D0-brane in IIA Wound: its rest energy in the parent description
is $1/\gs\ls$, which is finite and equal to $1/\Gs\Ls$
in the Wound limit. By transverse T-duality, one obtains all other
transverse D$p$-branes, and their tensions are found to be
\be \label{woundDpperp}
T^{\perp}_{Dp}= {1\over (2\pi)^{p}\Gs\Ls^{p+1}}~,
\ee
which unlike (\ref{woundDp}) agrees with the naive expectation.

Excitations of such branes will of course be described by
open strings which end on them. It is interesting
to note, however, that due to the
familiar decoupling argument these open strings must necessarily
have positive winding number along $x^{1}$.
So these are not standard D-branes--- in particular, the usual
massless modes associated with quantum fluctuations of their
positions are missing!
As a matter of fact,
the ground state configuration for these branes
includes at least one F-string attached to them,
since their total winding number $w$ is as always T$_1$-dual to the
(necessarily positive) longitudinal momentum of the
longitudinally wrapped DLCQ brane, $p'_{-}=w/R$.
In the parent type II string theory, the total
energy of the combined brane-string system is thus divergent, but
becomes finite after the usual subtraction.

Notice that one can also consider states with more than
one type of brane present. For instance, a DLCQ IIA state with a D0- and a
(transverse) D4-brane
is mapped onto a IIB Wound state which includes both a D1-brane
and a D5-brane (with the electric field on both branes becoming
critical simultaneously).
We emphasize that
these two objects do not give rise to two decoupled
theories: they can interact through the exchange of
wound closed strings. We leave a study of these interactions
to future work.

\section{Wound IIA/B Theory on a Transverse Torus}

\label{ncosontorussec}

Since the Wound IIA/B theory is  T$_{1}$-dual to DLCQ IIB/A,
it is a part of the well-known duality web
of  DLCQ IIA/M-theory 
\cite{bfss,susskind,seiberg,sen}.
In this section we examine this web of dual descriptions for 
Wound/DLCQ IIA/B
theory compactified
on a transverse $\mathbf{T}^{p-1}$, for $1\leq p\leq 5$. 
The parameters of the Wound and DLCQ
theories are in all cases related 
through the T-duality formulas obtained in Section \ref{dlcqsec}: 
\begin{equation}
\ensuremath{\tilde{R}_{-}}={\frac{\ensuremath{L_s}^{2}}{R_{1}}},\qquad 
\tilde{\ensuremath{g_s}}=\ensuremath{G_s}{\Ls\over R_{1}},\qquad \tilde{%
\ensuremath{l_s}}=\ensuremath{L_s}~.  \label{ncosvsdlcq}
\end{equation}
It is also of interest to examine the effect of S- and transverse T-duality,
to incorporate in the discussion the known S-duals of the various NCOS
theories \cite{om,kt}, and the Seiberg argument \cite{seiberg,sen} for DLCQ
M-theory on a transverse $\mathbf{T}^{p}$. The knowledge gained in Section 
\ref{ncossec} will allow us to make some new inferences about these other
descriptions. The situation for $p=1,3,5$ is summarized in Fig.~3, while
Fig.~4 displays the cases $p=2,4$. Of course, the theories encountered at
different values of $p$ are all connected through transverse T-dualities
followed by decompactification of some directions. As we will also
emphasize, the various
NCOS theories simply correspond to different states in the
universal Wound IIA/B theory. 

We would like to single out a few of the theories in this duality web, and
write down the explicit dictionary between their parameters and those of the
Wound theory. Which theories are interesting depends on the value of $p$, so
we will discuss each case separately. The case $p=1$ has already been
discussed at length in Section \ref{dlcqsec}, so we will omit it here. For
the other cases, we will only discuss theories which are `good' dual
descriptions of the $\epsilon\to 0$ physics, in the sense that the relevant
parameters are finite 
in this limit. The dual DLCQ IIA, DLCQ IIB and DLCQ M theories are good
descriptions for all values of $p$, but we will not write down their
parameters explicitly since the relevant formulas are essentially identical
to those given in Section \ref{dlcqsec}. In the discussion to follow we will
denote the coordinate radii of the transverse cycles of the torus by $r_{i}$%
, $i=2,\ldots,p-1$. These are held fixed in the wound limit (recall that the
effective Wound metric is simply $G_{\mu\nu}=\eta_{\mu\nu}$), which means
that the corresponding proper radii $R_{i}=r_{i}\sqrt{\epsilon}$ in the
parent type II description shrink to zero size in the limit (they are,
however, fixed in units of $\ensuremath{l_s}$).

\begin{figure}[htb]
\begin{center}
\begin{picture}(150,45) {\small
\put(0,40){\begin{minipage}[t]{4cm} \begin{center}
Wound IIB \\
N F1:1 \\
K Dp:1$\cdot\cdot$p
\end{center}\end{minipage}}
\put(0,10){\begin{minipage}[t]{4cm} \begin{center}
$\overline{\mbox{IIB}}$
on small $\mathbf{T}^{p-1}$\\
N D1:1 \\
K F1/D3/NS5:1$\cdot\cdot$p
\end{center}\end{minipage}}
\put(28,-20){\begin{minipage}[t]{4cm} \begin{center}
$\widehat{\mbox{IIB}}$ \\
N Dp:1$\cdot\cdot$p \\
K F1:1/D1:1/NS5:1$\cdot\cdot$5
\end{center}\end{minipage}}
\put(55,40){\begin{minipage}[t]{4cm} \begin{center}
IIA on small $\mathbf{T}^{p}$\\
N $P_{1}$ \\
K D(p-1):2$\cdot\cdot$p
\end{center}\end{minipage}}
\put(55,10){\begin{minipage}[t]{4cm} \begin{center}
$\overline{\mbox{IIA}}$ on small $\mathbf{T}^{p}$\\
N D0-branes\\
K $P_{1}$/D2:23/NS5:1$\cdot\cdot$5
\end{center}\end{minipage}}
\put(110,40){\begin{minipage}[t]{4cm} \begin{center}
DLCQ IIA \\
N \Pm \\
K D(p-1):2$\cdot\cdot$p
\end{center}\end{minipage}}
\put(110,10){\begin{minipage}[t]{4cm} \begin{center}
DLCQ M \\
N \Pm \\
K $P'_{10}$/M2:23/M5:2$\cdot\cdot$5
\end{center}\end{minipage}}
\put(30,-12){\vector(-1,1){6}}
\put(30,-12){\vector(1,-1){6}}
\put(31,-10){T$_{2\cdot\cdot p}$}
\put(47,41){\vector(-1,0){8}}
\put(47,41){\vector(+1,0){8}}
\put(45,44){T$_{1}$}
\put(105,41){\vector(-1,0){8}}
\put(105,41){\vector(+1,0){8}}
\put(103,44){$\beta_{1}$}
\put(47,12){\vector(-1,0){8}}
\put(47,12){\vector(+1,0){8}}
\put(45,15){T$_{1}$}
\put(105,12){\vector(-1,0){8}}
\put(105,12){\vector(+1,0){8}}
\put(103,15){$\beta_{1}$}
\put(130,21){\vector(0,-1){4}}
\put(130,21){\vector(0,+1){4}}
\put(133,20){S}
\put(73,21){\vector(0,-1){4}}
\put(73,21){\vector(0,+1){4}}
\put(75,20){10-1 flip}
\put(18,21){\vector(0,-1){4}}
\put(18,21){\vector(0,+1){4}}
\put(21,20){S} }
\end{picture}
\end{center}
\par
\vspace*{2.5cm}
\caption{{\protect\small A portion of the duality web for Type II Wound/DLCQ
theories on a transverse $\mathbf{T}^{p-1}$ for $p=1,3,5$, including the
images of $K$ Wound D$p$-branes and $N$ units of F-string winding in the
various descriptions. As explained in Section \ref{ncossec}, $N$ must be
strictly positive, but $K$ is arbitrary. The Wound IIB setup in the
upper-left corner of the diagram has hitherto been known as $(p+1)$%
-dimensional NCOS theory. Non-vanishing-size compactifications are not
mentioned. $\protect\beta_{1}$ denotes a boost along $x^{1}$ together with a
change of units. $P_{i}$ stands for Kaluza-Klein units of momentum along $%
x^{i}$. The notation Xq:$i_{1}\cdot\cdot i_{q-1}$ indicates an X$p$-brane
wrapping the $i_{1},\ldots,i_{q-1}$ cycles of the torus. Triple $K$-entries
apply respectively to the cases $p=1,3,5$. Seiberg's derivation of a
non-perturbative (Matrix) formulation for DLCQ IIA on a transverse $\mathbf{T%
}^{p-1}$ (DLCQ M on a transverse $\mathbf{T}^{p}$) follows the horizontal
arrows in the top (middle) line of the diagram, and then proceeds down and
diagonally to arrive at the $\widehat{\mbox{IIB}}$ theory. See text for
further discussion.}}
\end{figure}
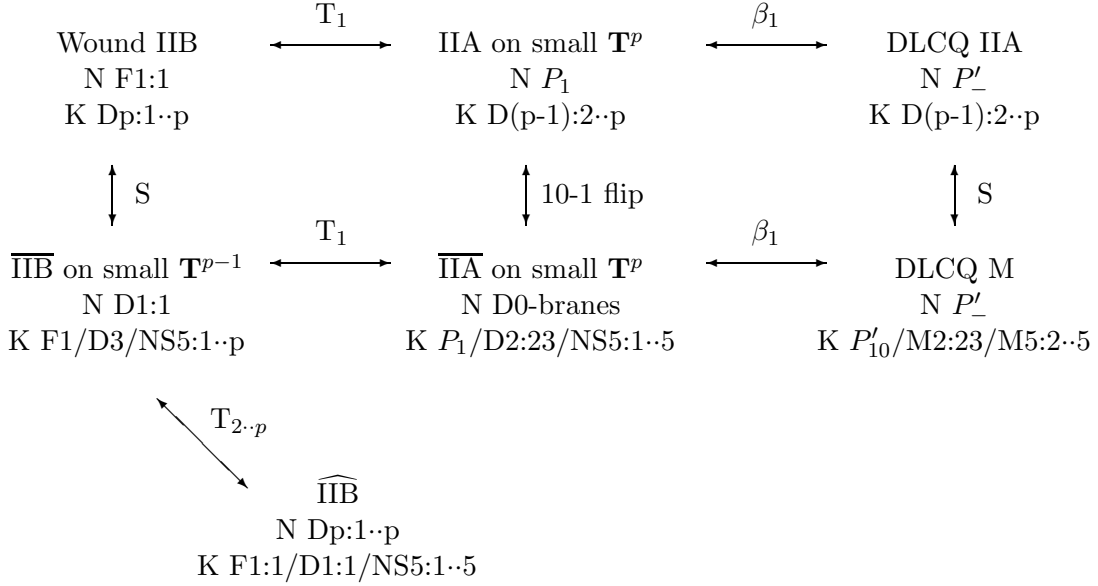

\subsection{$p=3$}

\label{3sec}

In the D3-brane case, the parameters of the theory termed $\overline{%
\mbox{IIB}}$ in Fig.~3 are related to those of the S-dual Wound theory
through 
\begin{equation}
\bar{\ensuremath{g_s}}=\ensuremath{G_s}^{-1}\sqrt{\epsilon }\rightarrow
0,\quad \bar{\ensuremath{l_s}}=\sqrt{\ensuremath{G_s}}\ensuremath{L_s}%
\epsilon ^{1/4}\rightarrow 0,\quad \bar{R}_{1}=R_{1},\quad \bar{R}%
_{2,3}=r_{2,3}\sqrt{\epsilon }\rightarrow 0.  \label{ncym}
\end{equation}
As seen in the figure, a collection of $K$ Wound D3-branes with $N$ units of
F-string winding are mapped onto $K$ $\overline{\mbox{IIB}}$ D3-branes with $%
N$ units of D-string winding. The theory on the three-branes is thus a $U(K)$
$(3+1)$-dimensional gauge theory with $N$ units of $F_{23}$-flux. As
explained in \cite{ncos}, the scaling of $\bar{\ensuremath{g_s}}$ and $\bar{%
\ensuremath{l_s}}$ seen in (\ref{ncym}) is precisely the Seiberg-Witten
scaling \cite{sw} which yields non-commutative super-Yang-Mills theory
(NCYM). Just like in the Wound description, the proper radii of the
non-commutative $x^{2},x^{3}$ directions are fixed when measured in the open
string metric relevant to the NCYM theory.

As seen in (\ref{ncym}), the proper radii $R_{2,3}$ (measured in the
underlying $\overline{\mbox{IIB}}$ metric) shrink to zero size in the
Wound/DLCQ limit, so it is natural to take the final step in the Seiberg
argument \cite{seiberg,sen}, following the diagonal arrow in Fig.~3 (T$_{23}$%
-duality) to arrive at a $\widehat{\mbox{IIB}}$ theory defined on a
finite-size three-torus, with parameters 
\begin{equation}
\hat{\ensuremath{g_s}}={\frac{\ensuremath{L_s}^{2}}{r_{2}r_{3}}},\quad \hat{%
\ensuremath{l_s}}=\sqrt{\ensuremath{G_s}}\ensuremath{L_s}\epsilon
^{1/4}\rightarrow 0,\quad \hat{R}_{1}=R_{1},\quad \hat{R}_{2,3}={\frac{%
\ensuremath{G_s}\ensuremath{L_s}^{2}}{r_{2,3}}}.  \label{matrix3}
\end{equation}
The Wound IIB D3- and F1-branes now become $K$ D1-branes and $N$ D3-branes,
so we are dealing with $(3+1)$-dimensional $U(N)$ (ordinary) SYM with $K$
units of $F_{23}$-flux. The Yang-Mills coupling for the theory is 
\begin{equation}
\hat{g}_{YM}^{2}=2\pi \hat{\ensuremath{g_s}}={\frac{2\pi \ensuremath{L_s}^{2}%
}{r_{2}r_{3}}}  \label{matrix3coupling}
\end{equation}

Clearly, there is nothing that prevents us from setting $K=0$, neither in
the case of $\overline{\mbox{IIB}}$, nor in the case of 
$\widehat{\mbox{IIB}}$.
In the former case 
we notice that D1-branes with strictly
positive (D-string) winding along $x^{1}$ can be emitted at a
finite energy cost into the bulk, 
where they can be studied on their own. 
It is thus sensible to consider this theory even in the absence of D3-branes
(i.e., $K=0$): it is just the S-dual of the Wound IIB theory without any
D-branes. We will refer to this as the Wound D-string theory. Notice that
these two theories are really distinct: the strong/weak-coupling
self-duality of the parent IIB theory is \emph{not} inherited by the Wound
IIB theory, because the wound limit scales $\ensuremath{g_s}\rightarrow
\infty $. 

Turning to the case of $\widehat{\mbox{IIB}}$, where the $K$
D3-branes have turned into $K$ D-strings and the $N$ D1-strings have turned
into $N$ D3-branes, the possibility for the wound D3-branes to emit a wound
string into the bulk becomes a rather mundane symmetry breaking process. 
This is well-understood in the Matrix context, and is the obvious analog of
the $1+1$ SYM breaking studied in \cite{km} and reviewed in Section \ref
{dlcqsec}. In this description it is therefore also evident that there is
nothing peculiar about the case $K=0$ (where all the wound strings have been
emitted). In fact, this is nothing but the Matrix model conjectured to
provide a non-perturbative description of DLCQ M-theory on a transverse $%
\mathbf{T}^{3}$ \cite{bfss,wati}.

One curious aspect of (\ref{matrix3}) and (\ref{matrix3coupling}) is the
fact that the coupling constant in this description is independent of the
Wound coupling. Notice, however, that this does not lead one to the
contradiction of having two manifestly distinct weakly-coupled descriptions
of the same underlying physical system: simultaneously requiring that $%
\ensuremath{G_s}\ll 1$ and $\hat{g}_{YM}^{2}\ll 1$ can be seen from (\ref
{matrix3}) to imply that $r_{2,3}\ll \ensuremath{L_s}$, which means the
Wound theory is not really amenable to a 
direct perturbative analysis. 

\subsection{$p=5$}

\label{5sec}

In the case of D5-branes, the $\overline{\mbox{IIB}}$ theory parameters are
again given by (\ref{ncym}), except that there are two more transverse
directions, $x^{4},x^{5}$. As noted in Fig.~3, the $K$ Wound D5-branes and $%
N $ F-strings are now mapped onto $K$ $\overline{\mbox{IIB}}$ NS5-branes and 
$N $ units of D-string winding. The latter manifest themselves as $N$ units
of $\mathcal{F}_{01}$-flux in the NS5 worldvolume, where $\mathcal{F}$ is
the two-form field strength (which is S-dual to $F$ on the original
D5-branes). As explained in \cite{om,harmark2}, 
$\mathcal{F}$ (or equivalently, the Ramond-Ramond $C_{01}$ potential)
becomes critical in the Wound limit, which together with the scaling (\ref
{ncym}) yields the so-called OD1 theory, whose excitations are open
D-strings ending on the NS5-branes. Interactions of these open D-strings are
governed by the coupling \cite{om}  
\begin{equation}
G_{D1}={\frac{\hat{\ensuremath{g_s}}}{\sqrt{\epsilon }}}={\frac{1}{%
\ensuremath{G_s}}}~,  \label{OD1coupling}
\end{equation}
and the theory has a length parameter 
\begin{equation}
L_{D1}={\frac{\hat{\ensuremath{l_s}}}{\epsilon ^{1/4}}}=\sqrt{%
\ensuremath{G_s}}\ensuremath{L_s}~.  \label{OD1length}
\end{equation}

Again, it is natural to take the next step in the Seiberg argument \cite
{seiberg,sen}, carrying out a T$_{2345}$-duality transformation to arrive at
a $\widehat{\mbox{IIB}}$ theory defined on a finite-size five-torus, with
parameters 
\begin{equation}  \label{OD5}
\hat{\ensuremath{g_s}}={\frac{1}{\ensuremath{G_s}}}{\frac{\ensuremath{L_s}%
^{4}}{r_{2}r_{3}r_{4}r_{5}}}\epsilon^{-1/2}\to\infty, \quad \hat{%
\ensuremath{l_s}}=\sqrt{\ensuremath{G_s}}\ensuremath{L_s}\epsilon^{1/4}\to
0, \quad \hat{R}_{1}=R_{1}, \quad \hat{R}_{i}={\frac{\ensuremath{G_s}%
\ensuremath{L_s}^{2}}{r_{i}}},
\end{equation}
in the presence of $K$ NS5-branes and $N$ D5-branes. This is the OD5 theory
of \cite{om,harmark2}, 
whose excitations are described by open D5-branes ending on the NS5
worldvolume interacting with strength 
\begin{equation}  \label{OD5coupling}
G_{D5}={\frac{1}{\ensuremath{G_s}}}{\frac{\ensuremath{L_s}^{4}}{%
r_{2}r_{3}r_{4}r_{5}}}~,
\end{equation}
and length parameter 
\begin{equation}  \label{OD5length}
L_{D5}=\sqrt{\ensuremath{G_s}}\ensuremath{L_s}~.
\end{equation}

Just as in the case of $p=1$ and $p=3$ we are free to set $K=0$. 
For $\overline{\mbox{IIB}}$, this implies that one
can take the `OD1' limit even in the absence of NS5-branes.
This yields a theory of D1-branes with strictly positive winding which is
identical to the Wound D-string theory encountered in the previous
subsection. That this theory is truly S-dual to the (IIB) Wound F-string
theory is made explicit by the relation (\ref{OD1coupling}). The $K=0$ state
in this theory does not have an open D-string sector--- the role of these
open strings in the theory is merely to describe excitations of NS5-branes
when they happen to be present. It should also be noted that a
generalization of the decoupling argument of Section \ref{ncossec} implies
that the presence of a near-critical $C_{01}$-field is not strictly
necessary in the Wound D-string (OD1) limit, although of course this field
is a useful tool to implement the energy subtraction. It can be seen from
Fig.~3 that the Wound D-string theory is T$_{1}$-dual to DLCQ M-theory. This
relation is just the S-dual image of the Wound IIB $\leftrightarrow $ DLCQ
IIA T$_{1}$-duality. The mapping between the two theories converts D-string
winding into longitudinal momentum, and NS5-branes into M5-branes. 

Turning to $\widehat{\mbox{IIB}}$ with $K=0$, we infer that IIB string
theory on a five-torus has a limit (the `OD5' limit) in which the basic
degrees of freedom are wrapped D5-branes with strictly positive wrapping
number. Extending our earlier nomenclature, it is natural to refer to this
as the Wrapped D5-brane theory.

Let us pause here to note that one can similarly define Wrapped D$p$-brane
theories for $p=2,3,4$, which are all related to one another and to the
Wrapped D1/D5-brane theories by T-dualities along directions $2345$. It
should be clear that for each value of $p$, the OD$p$ theory constructed in 
\cite{om,harmark2} is simply a truncation of the Wrapped D$p$-brane theory
to the set of states containing NS5-branes which extend along the `Wrapped'
directions. 
The case $p=0$ is special: here the names `OD0' or `Wrapped' are evidently
not appropriate. The restriction that the `Wrapped' limit induces in this
case is simply that the D0-brane charge of all states be strictly positive,
so as pointed out in \cite{om} this is nothing but the original Matrix
description of DLCQ M-theory \cite{bfss,susskind}. We had already noted
above that this theory is T$_{1}$-dual to the Wound D-string theory. 

Let us now return to the $p=5$ case. Since the $\widehat{\mbox{IIB}}$
coupling (\ref{OD5}) diverges, Seiberg's argument \cite{seiberg,sen} for
this case requires a final S-duality transformation (not shown in the
figure) to obtain a weakly-coupled IIB$^{\prime }$ theory with 
\begin{equation}
\ensuremath{g_s}^{\prime }=\ensuremath{G_s}{r_{2}r_{3}r_{4}r_{5}\over\Ls^{4}}%
\epsilon ^{1/2}\rightarrow 0,\quad \ensuremath{l_s}^{\prime }={\frac{%
\ensuremath{L_s}^{3}}{\sqrt{r_{2}r_{3}r_{4}r_{5}}}},\quad R_{1}^{\prime
}=R_{1},\quad R_{i}^{\prime }={\frac{\ensuremath{G_s}\ensuremath{L_s}^{2}}{%
r_{i}}},  \label{matrix5}
\end{equation}
with $K$ D5-branes and $N$ NS5-branes. The scaling $\ensuremath{g_s}^{\prime
}\rightarrow 0$ at fixed $\ensuremath{l_s}^{\prime }$ would appear to yield
the decoupled $(1,1)$ Little String Theory (LST) on the NS5-brane
worldvolume \cite{lst}. However, as has been emphasized in \cite{om}, for $%
K\neq 0$ the presence of the D5-branes implies that the IIB$^{\prime }$ is
in fact strongly-coupled. There is no such obstruction in the case $K=0$, so
we conclude that Wound IIB theory with no D5-branes is (S$\cdot $T$%
_{2345}\cdot $S)-dual 
to $(1,1)$ LST.

\subsection{$p=2$}

\label{2sec}

As indicated in Fig.~4, Wound IIA on $\mathbf{T}^{2}$ can be lifted to
M-theory on $\mathbf{T}^{3}$, with parameters 
\begin{equation}  \label{M}
\ensuremath{l_P}=\ensuremath{G_s}^{1/3}\ensuremath{L_s}\epsilon^{1/3}\to
0~,\quad R_{1}=R_{1}, \quad R_{2}=r_{2}\sqrt{\epsilon}\to 0, \quad R_{10}=%
\ensuremath{G_s}\ensuremath{L_s}~.
\end{equation}
Under the lift, the $K$ Wound D2-branes and $N$ F-strings become an M2-brane
bound state with wrapping number $N>0$ on the $x^{1}$-$x^{10}$ torus 
\cite{om}.
In other words, the Wound limit of IIA maps to a limit (\ref{M}) of M-theory
which involves singling out two compact directions (in this case $1$ and $10$%
) and scaling $\ensuremath{l_P}\to 0$ in such a way that only objects with
strictly positive M2 wrapping number on the 1-10 torus remain in the
spectrum. Leaving $K$ arbitrary means that M2-branes are free to also wrap
around the $x^{2}$ circle; $K=0$ is the special case where the M2-branes
extend solely along $x^{1}$ and $x^{10}$. Extending our previous
terminology, it is appropriate to refer to this as the Wrapped M2-brane
theory. We will return to it in the next subsection.

Regarding $x^{2}$ as the M-theory circle, one can descend (following the
dotted arrow in Fig.~4) to a ten-dimensional theory denoted as $\widehat{%
\mbox{IIA}}$ in Fig.~4, with 
\begin{equation}  \label{matrix2}
\hat{\ensuremath{g_s}}= \ensuremath{G_s}^{-1/2}\left({\frac{r_{2}}{%
\ensuremath{L_s}}}\right)^{3/2}\epsilon^{1/4}\to 0, \quad \hat{%
\ensuremath{l_s}}=\sqrt{\frac{\ensuremath{G_s}\ensuremath{L_s}^{3}}{r_{2}}}%
\epsilon^{1/4}\to 0, \quad \hat{R}_{1}=R_{1}, \quad \hat{R}_{2}=%
\ensuremath{G_s}\ensuremath{L_s},
\end{equation}
$N$ D2-branes, and $K$ units of F-string winding. This is $(2+1)$%
-dimensional $U(N)$ SYM with $K$ units of $F_{02}$-flux. This is the theory
which Seiberg's argument \cite{seiberg,sen} puts forth as the
non-perturbative (Matrix) description of the system in question \cite
{bfss,wati}. 
Its coupling constant is 
\begin{equation}  \label{matrix2coupling}
g_{YM}^{2}={\frac{\hat{\ensuremath{g_s}}}{\hat{\ensuremath{l_s}}}}={\frac{%
R_{1}^{2}}{\ensuremath{L_s}^{3}}}~.
\end{equation}
It is again obvious from this perspective that one is free to set $K=0$.
Just like in the $p=3$ case, it is curious to note that this is independent
of the Wound coupling \ensuremath{G_s}, but again it should be emphasized
that this does not imply the existence of some regime where the two
descriptions can be examined perturbatively at the same time. 

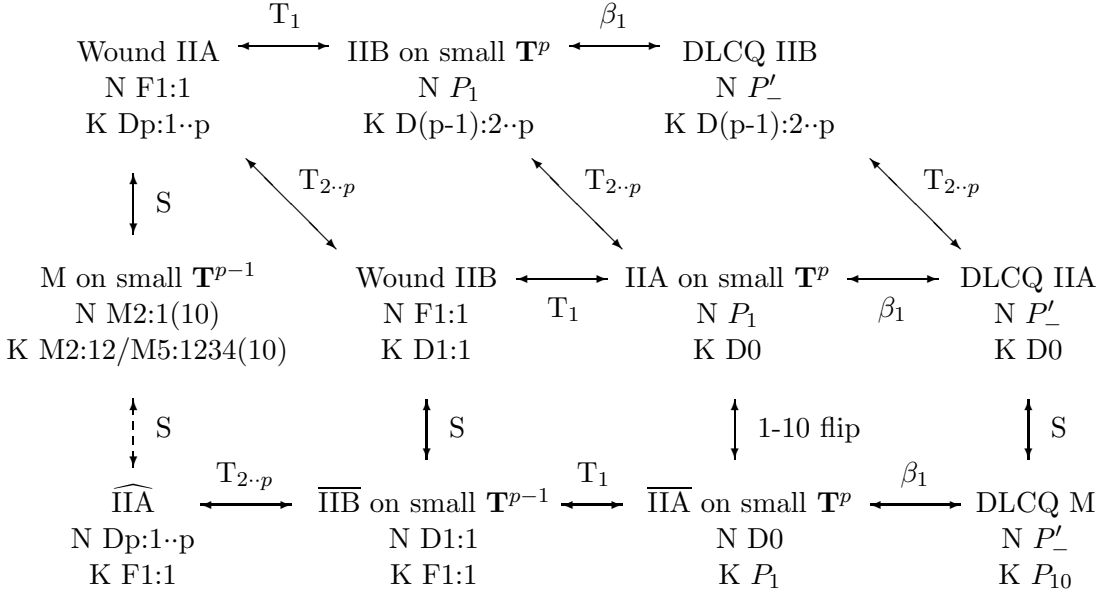
\begin{figure}[tb]
\begin{center}
\vspace{0.5cm} 
\begin{picture}(150,75) {\small
\put(0,70){\begin{minipage}[t]{4cm} \begin{center}
Wound IIA \\
N F1:1 \\
K Dp:1$\cdot\cdot$p
\end{center}\end{minipage}}
\put(40,70){\begin{minipage}[t]{4cm} \begin{center}
$\mbox{IIB}$ on small $\mathbf{T}^{p}$\\
N $P_{1}$ \\
K D(p-1):2$\cdot\cdot$p
\end{center}\end{minipage}}
\put(80,70){\begin{minipage}[t]{4cm} \begin{center}
DLCQ IIB \\
N \Pm \\
K D(p-1):2$\cdot\cdot$p
\end{center}\end{minipage}}
\put(0,40){\begin{minipage}[t]{4cm} \begin{center}
M on small $\mathbf{T}^{p-1}$\\
N M2:1(10) \\
K M2:12/M5:1234(10)
\end{center}\end{minipage}}
\put(37,40){\begin{minipage}[t]{4cm} \begin{center}
Wound IIB \\
N F1:1 \\
K D1:1
\end{center}\end{minipage}}
\put(77,40){\begin{minipage}[t]{4cm} \begin{center}
IIA on small $\mathbf{T}^{p}$ \\
N $P_{1}$ \\
K D0
\end{center}\end{minipage}}
\put(117,40){\begin{minipage}[t]{4cm} \begin{center}
DLCQ IIA \\
N \Pm \\
K D0
\end{center}\end{minipage}}
\put(-2,10){\begin{minipage}[t]{4cm} \begin{center}
$\widehat{\mbox{IIA}}$\\
N Dp:1$\cdot\cdot$p \\
K F1:1
\end{center}\end{minipage}}
\put(38,10){\begin{minipage}[t]{4cm} \begin{center}
$\overline{\mbox{IIB}}$ on small $\mathbf{T}^{p-1}$\\
N D1:1\\
K F1:1
\end{center}\end{minipage}}
\put(80,10){\begin{minipage}[t]{4cm} \begin{center}
$\overline{\mbox{IIA}}$ on small $\mathbf{T}^{p}$ \\
N D0 \\
K $P_{1}$
\end{center}\end{minipage}}
\put(118,10){\begin{minipage}[t]{4cm} \begin{center}
DLCQ M \\
N \Pm \\
K $P_{10}$
\end{center}\end{minipage}}
\put(38,72){\vector(-1,0){6}}
\put(38,72){\vector(+1,0){6}}
\put(36,75){T$_{1}$}
\put(82,72){\vector(-1,0){6}}
\put(82,72){\vector(+1,0){6}}
\put(80,75){$\beta_{1}$}
\put(18,51){\vector(0,-1){4}}
\put(18,51){\vector(0,+1){4}}
\put(21,50){S}
\put(39,51){\vector(-1,1){6}}
\put(39,51){\vector(1,-1){6}}
\put(40,53){T$_{2\cdot\cdot p}$}
\put(77,51){\vector(-1,1){6}}
\put(77,51){\vector(1,-1){6}}
\put(78,53){T$_{2\cdot\cdot p}$}
\put(122,51){\vector(-1,1){6}}
\put(122,51){\vector(1,-1){6}}
\put(123,53){T$_{2\cdot\cdot p}$}
\put(75,41){\vector(-1,0){6}}
\put(75,41){\vector(+1,0){6}}
\put(73,36){T$_{1}$}
\put(119,41){\vector(-1,0){6}}
\put(119,41){\vector(+1,0){6}}
\put(117,36){$\beta_{1}$}
\put(18,19){\line(0,1){1}}
\put(18,21){\line(0,1){1}}
\put(18,18){\vector(0,-1){2}}
\put(18,23){\vector(0,+1){2}}
\put(21,20){S}
\put(57,21){\vector(0,-1){4}}
\put(57,21){\vector(0,+1){4}}
\put(60,20){S}
\put(98,21){\vector(0,-1){4}}
\put(98,21){\vector(0,+1){4}}
\put(101,20){1-10 flip}
\put(137,21){\vector(0,-1){4}}
\put(137,21){\vector(0,+1){4}}
\put(140,20){S}
\put(33,11){\vector(-1,0){6}}
\put(33,11){\vector(+1,0){6}}
\put(29,14){T$_{2\cdot\cdot p}$}
\put(79,11){\vector(-1,0){4}}
\put(79,11){\vector(+1,0){4}}
\put(77,14){T$_{1}$}
\put(122,11){\vector(-1,0){6}}
\put(122,11){\vector(+1,0){6}}
\put(120,14){$\beta_{1}$}
}
\end{picture}
\end{center}
\par
\vspace*{-0.5cm}
\caption{{\protect\small Same as Fig. 2, for $p=2,4$. 
The two figures are related through transverse
T-duality. Again, compact directions are not mentioned unless
they have vanishing size in the relevant metric.
Double entries for the object with multiplicity $K$ apply
respectively to the $p=2,4$ cases. The dotted arrow connecting M-theory to 
$\widehat{\mbox{IIA}}$ applies only in the $p=2$ case. The Wound IIA setup in
the upper-left corner has hitherto been known as $p+1$ NCOS theory.
Seiberg's derivation of a non-perturbative (Matrix) description of DLCQ
M-theory on $\mathbf{T}^{p}$ proceeds along the bottom line of the figure.}}
\end{figure}

\subsection{$p=4$}

The direct lift of Wound IIA to eleven dimensions takes us to M-theory with
parameters (\ref{M}), $R_{3,4}=r_{3,4}\sqrt{\epsilon }$, in the presence of $%
N$ M2-branes and $K$ M5-branes. When bound to the fivebranes, the membranes
appear as $N$ units of $H_{01(10)}$-flux, where $H$ is the (self-dual) 
three-form field strength on the fivebrane worldvolume. 
This field (or equivalently, the bulk $A_{01(10)}$ gauge field) becomes
critical in the Wound/DLCQ limit, 
yielding a framework known as OM theory \cite{om,bbss2}. 
Reasoning like before, we
see that the possibility of setting $K=0$ means that it is possible to
define an `OM' limit of M-theory even if there are no M5-branes present, to
obtain a theory which contains M2-branes with strictly positive wrapping
number on the $x^{1}$-$x^{10}$ torus. This is of course the Wrapped M2-brane
theory discussed in the previous subsection. M5-branes remain in the
spectrum of this theory only if they carry a positive M2 1-10 wrapping
number. When they are present, their excitations include open M2-branes---
this is the standard OM theory setup.

We should emphasize that, despite the fact that $\ensuremath{l_P}\to 0$, the
rescaling of the transverse directions seen in (\ref{M}) implies that the
Wrapped M2 theory is not just the conformal $SO(8)$-invariant theory on the
M2-brane worldvolume--- rather, it is a limit of the full eleven-dimensional
M-theory. States in this theory can contain several interacting clumps of
M2-branes with distinct (positive) $1-10$ wrapping number, or even
M5-branes, as we have seen above. 

Following Seiberg's procedure \cite{seiberg,sen} one arrives in this case at
the theory labelled $\widehat{\mbox{IIA}}$ in Fig.~4, 
only to realize that its coupling constant diverges, signalling the need to
employ an $\widehat{\mbox{M}}$-theory description (not shown in the figure)
with 
\begin{equation}  \label{matrix4}
\hat{\ensuremath{l_P}}=\ensuremath{G_s}^{2/3}{\frac{\ensuremath{L_s}^{2}}{%
(r_{2}r_{3}r_{4})^{1/3}}}\epsilon^{1/6}\to 0, \quad \hat{R}_{1}=R_{1}, \quad 
\hat{R}_{i}=\ensuremath{G_s}{\Ls^{2}\over r_{j}r_{k}}, \quad \hat{R}_{10}=%
\ensuremath{G_s}\ensuremath{L_s}~.
\end{equation}
In the third equation $i,j,k$ are understood to take different values, so
for example $\hat{R}_{2}=\ensuremath{G_s}\ensuremath{L_s}^{2}/r_{3}r_{4}$.
In this picture there are $N$ M5-branes and $K$ M2-branes present, and the
limit $\ensuremath{l_P}\to0$ yields the decoupled $(2,0)$ theory on the
M5-brane worldvolume. This is the theory conjectured to capture the physics
of DLCQ M-theory on $\mathbf{T}^{4}$ at the non-perturbative level \cite
{rozali,brs}. The $K$ M2-branes are realized as $H_{01(10)}$-flux 
in the $(2,0)$ theory, and as in all other cases it is clear that one is
free to set $K=0$.

\section{Conclusions} \label{conclusions}

In this paper we have analyzed the T-duality relation
between the DLCQ limit of IIA/B
theories and the NCOS theories \cite{sst2,ncos,km}. In the process, we
have discovered that it is meaningful to consider an `NCOS' limit
of the IIA/B theory without branes. We
compactify the IIA/B theory on a circle of radius $R$ and consider a sector
with definite, conserved, total winding number $W$. The low energy limit 
$l_{s}\rightarrow 0$ can be taken keeping $g_{s}l_{s}$ and $R$ fixed, and
turns out to be exactly T-dual to the limit that defines the DLCQ IIB/A
with $W$
units of momentum along the lightlike direction. 
One finds that the energy
of a string with winding number $w$ has a contribution $|w|R/l_{s}^{2}$
which diverges in the limit. If we restrict attention to the 
subsector where all winding numbers are strictly positive 
($|w|=w>0$), this divergent
contribution can be subtracted, since the total winding  
$W=\sum w_{i}=\sum |w_{i}|$ 
is conserved in all interactions. As we have discussed in Section 
\ref{clstrsec},
the subtraction is equivalent to considering the theory in the presence of
a critical $B_{01}$-field, but since this field can be gauged away the
physics cannot change. The T-dual procedure in the DLCQ limit is to boost
along the circle, which also does not change the physics. The rules of
T-duality show that boosting and turning on
a $B_{01}$-field are precisely equivalent.

We call the resulting theory the Wound IIA/B theory, 
since it is characterized by the fact that all objects in it  
carry strictly positive F-string winding. At low energy the
loop-counting parameter in the theory is $g_{s}l_{s}/R$, as we have checked
in Section \ref{amplsec} by means of
a one-loop calculation. However, since in the IIB case the Wound theory
is in a certain sense (see below)
S-dual to $1+1$ $U(W)$ SYM with $g_{YM}\sim 1/(g_{s}l_{s})$ \cite{om,km},
which is strongly coupled in the
infrared, we expect the effective coupling at energies $E\gg 1/R$ to grow as 
$g_{s}l_{s}E$. For these energies the length scale $g_{s}l_{s}$ is the
meaningful parameter of the theory. The scattering amplitudes in the Wound
IIA/B theory are similar to those in the usual string theory, even though
the momentum energy relation is non-relativistic with mass and energy
separately conserved (in this context winding number is interpreted as a
mass). This is clear since the theory is perturbatively equivalent (T-dual)
to the standard DLCQ, replacing winding number by momentum in the compact
direction. 

In Section \ref{dbranesec} we have considered the Wound 
theory in the presence of D-branes. We have seen there
that in order for a brane to remain in the spectrum, it must carry an
electric flux along the Wound circle, or, equivalently,  
a strictly positive fundamental-string winding
number. In that case the limit for the theory on the brane reduces to the
corresponding NCOS theory, and the Wound theory emerges as the 
framework that unifies these setups, which up to now had been regarded
as isolated theories. 
T-duality implies that NCOS theory is also the
appropriate way to introduce D-branes in DLCQ IIA. For example, as shown in
Section \ref{ncostsec}, the non-commutativity in the 
DLCQ picture arises from the fact that left- and right-moving
momenta are different, which, as a consequence, gives a phase
when two open string vertices are interchanged. 
In the 
decompactification limit $R\to\infty$ the NCOS theories are obtained
as decoupled theories on the branes.

In Section \ref{ncosontorussec} we studied the Wound theory 
compactified on transverse tori. By use of diverse dualities, we were 
naturally led to various Wrapped $p$-Brane theories, which are 
obtained as limits 
of IIA/B/M-theory that leave in the spectrum only those objects 
which carry strictly positive brane
wrapping number on a specific $p$-torus.
Just like the Wound F-string theory unifies the various NCOS setups,
these Wrapped Brane theories contain
the extensions of NCOS theories to open $p$-brane theories,
i.e., OM theory \cite{om,bbss2} and OD$p$ theories \cite{harmark2,om}.
These theories have already appeared in the process of deriving Matrix
models for DLCQ IIA/M-theory on a transverse torus \cite{seiberg,sen},
but our improved understanding allows us
to consider them as theories in their own right.

As discussed in Section \ref{2sec}, as long as there are (at least) two compact
dimensions, it is possible to take a limit of M-theory which truncates the
spectrum down to those objects carrying strictly positive M2 wrapping
number.
This is the Wrapped M2-brane theory, which includes OM theory as a
particular set of states--- namely, those that contain M5-branes extended
along the `wrapped' directions. This M-theoretic construct is related to 
similar string-theoretic frameworks in the usual manner. If we
start with the Wrapped M2-brane theory and descend to ten dimensions by
identifying one of the `Wrapped' directions as the M-theory circle, we
obtain the Wound IIA theory. If, on the other hand, the M-theory circle is
transverse to the directions of M2 wrapping, the result is the Wrapped
D2-brane theory. 
Using transverse T-duality one can then reach the Wound IIB theory, and all
the other Wrapped Dp-brane theories. Additionally, starting with the Wound (IIB)
F-string and D-string theories, longitudinal T-duality brings us to DLCQ IIA
and DLCQ M-theory, respectively.
Finally, one is led to
propose the existence of the analogous theories of Wrapped
NS5-branes and M5-branes when seeking S-dual descriptions for all
of the Wrapped brane theories encountered so far.

We would like to emphasize that the Wrapped point of view 
is distinct from that of Matrix theory \cite{bfss,susskind}. 
When attempting to provide a non-perturbative formulation of
DLCQ IIA/M-theory on transverse tori of different dimensions,
the Matrix approach
focuses attention on models which are defined on spacetimes 
of different dimensionality \cite{seiberg,sen}.  The Wrapped $p$-brane
theories, on the other hand, are always defined on
a ten- or eleven-dimensional spacetime (with at least $p$
compact directions).  

To make this difference clearer, let us pose 
a specific question: what is the S-dual of the Wound IIB theory?
Since Wound IIB is defined as IIB in the limit (\ref{woundlim}), the 
answer is clear: S-dualizing one obtains a \emph{ten-dimensional}
theory which is defined as IIB in the limit
\be \label{woundD1lim}
\delta\to 0, \quad
\gs={1\over G_{s}}\sqrt{\delta}, \quad  
\ls=\sqrt{\Gs}\Ls\delta^{1/4}, \quad
h=\delta, \quad 
\mbox{with}\quad \Gs,\Ls,R\quad \mbox{fixed}.
\ee
This is true independently of whether the Wound theory is 
compactified on a transverse torus or not. Since the requirement that 
objects in the Wound theory carry strictly positive F-string winding 
is mapped to the condition that objects in the S-dual theory carry 
strictly positive D-string winding, it is reasonable to call this the 
Wound D-string theory. The relation to the Matrix approach is that, 
\emph{in the absence of a transverse compactification}, the limit 
(\ref{woundD1lim}) reduces the theory on the D1-brane worldvolume to 
$1+1$ SYM (i.e., Matrix String theory). If we disregard branes of 
infinite extent, 
then with only one compact direction there are no other possible branes 
in the Wound D-string theory\footnote{Except of course F-strings with 
adsorbed D-strings, which are easily accommodated in the $1+1$ SYM
description \cite{km}.}, so it can be effectively identified 
with $1+1$ SYM. 

Suppose now that we compactify the Wound IIB theory on,
say, a (transverse) two-torus. The S-dual Wound D-string theory
can then 
have D3-branes wrapped on the resulting (transverse + longitudinal) 
three-torus, as long as they carry positive D-string winding. This 
means there is a magnetic flux on the D3-brane worldvolume, and as 
noted in \cite{ncos} and
in Section \ref{3sec} of the present paper, 
the limit (\ref{woundD1lim}) yields the 
$(3+1)$-dimensional NCYM theory. Similarly, if we compactify on a 
transverse four-torus, there can be NS5-branes carrying D-string 
winding number, and the limit (\ref{woundD1lim}) defines what has been 
called the OD1 theory \cite{harmark2,om} (open D-strings describe 
excitations of the NS5-brane). The role of the Wound D-string theory 
as a unifying framework which can accommodate
all of these different degrees of freedom is thus clear. 
It is also evident that $1+1$ SYM cannot play this same role 
(at least not directly, and for finite $N$). 
The reason is that in the SYM 
description, the radii of the transverse circles are shrinking to 
zero size, due to the scaling of the transverse metric parameter $h$ 
built into (\ref{woundD1lim}). The Matrix approach \cite{seiberg,sen} 
therefore considers this description useless, and carries out further 
dualities to arrive at $(3+1)$-dimensional SYM and $(5+1)$-dimensional
$(1,1)$ LST, respectively.

Summarizing, the NCOS theories describe branes in the 
Wound IIB/A theory T-dual to the
full DLCQ IIA/B theory. Similarly,
OM theory and the ODp theories describe M5-branes and NS5-branes in
the Wrapped M2 and Wrapped Dp-brane theories--- theories which are 
U-dual to Wound IIA/B. It would be worthwhile to explore the possible 
additional implications that this unified perspective could have for 
the theories involved. 
In particular, the NCOS theories on a circle
provide a simple way to capture the physics of D-branes in the DLCQ limit of
string theory: through T-duality each brane is described by a
non-commutative open string theory. It would be interesting to
use this correspondence to try to gain a better understanding
of the physics of DLCQ.

\vspace*{1cm}
\noindent
{\textbf{Note Added:}} 
While this paper was being written,
the work \cite{hyun} appeared, which overlaps with our
Sections \ref{ncostsec} and \ref{ncosontorussec}.

\section{Acknowledgements}

AG would like to thank Ansar Fayyazuddin, 
Bo Sundborg, and especially Subir Mukho\-padhyay
for valuable conversations. We are grateful to F. Kristianson and
P. Rajan for useful discussions.
The work of UD and AG was supported by the Swedish Natural
Science Research Council (NFR).


\begin{thebibliography}{99}

    
\bibitem{cds}
A.~Connes, M.~R.~Douglas and A.~Schwarz,
``Noncommutative geometry and matrix theory: Compactification on tori,''
JHEP {\bf 9802}, 003 (1998),
\hepth{9711162}.
   
\bibitem{dh}
M.~R.~Douglas and C.~Hull,
``D-branes and the noncommutative torus,''
JHEP {\bf 9802}, 008 (1998),
\hepth{9711165}.

\bibitem{sw}
N.~Seiberg and E.~Witten,
``String theory and noncommutative geometry,''
JHEP {\bf 9909}, 032 (1999),
\hepth{9908142}.

\bibitem{sst2}
N.~Seiberg, L.~Susskind and N.~Toumbas,
``Strings in background electric field, space/time
non-commutativity and a new noncritical string theory,''
JHEP {\bf 0006}, 021 (2000),
\hepth{0005040}.

\bibitem{ncos}
R.~Gopakumar, J.~Maldacena, S.~Minwalla and A.~Strominger,
``S-duality and noncommutative gauge theory,''
JHEP {\bf 0006}, 036 (2000),
\hepth{0005048}.

\bibitem{om}
R.~Gopakumar, S.~Minwalla, N.~Seiberg and A.~Strominger,
``OM theory in diverse dimensions,''
\hepth{0006062}.

\bibitem{agm}
O.~Aharony, J.~Gomis and T.~Mehen,
``On theories with light-like noncommutativity,''
JHEP {\bf 0009}, 023 (2000)
\hepth/0006236.

\bibitem{sst1}
N.~Seiberg, L.~Susskind and N.~Toumbas,
``Space/time non-commutativity and causality,''
JHEP {\bf 0006}, 044 (2000),
\hepth{0005015}.

\bibitem{br}
J.~L.~Barbon and E.~Rabinovici,
``Stringy fuzziness as the custodian of time-space noncommutativity,''
Phys.\ Lett.\  {\bf B486}, 202 (2000),
\hepth{0005073}.

\bibitem{gm}
J.~Gomis and T.~Mehen,
``Space-time noncommutative field theories and unitarity,''
hep-th/0005129.


\bibitem{grs}
O.~J.~Ganor, G.~Rajesh and S.~Sethi,
``Duality and non-commutative gauge theory,''
\hepth{0005046}.

\bibitem{km}
I.~R.~Klebanov and J.~Maldacena,
``1+1 dimensional NCOS and its U(N) gauge theory dual,''
\hepth{0006085}.

\bibitem{hk}
C.~P.~Herzog and I.~R.~Klebanov,
``Stable massive states in 1+1 dimensional NCOS,''
\hepth{0009017}.

\bibitem{harmark1}
T.~Harmark,
``Supergravity and space-time non-commutative open string theory,''
JHEP {\bf 0007}, 043 (2000),
\hepth{0006023}.

\bibitem{sahakian}
V.~Sahakian,
``The phases of 2-D NCOS,''
JHEP {\bf 0009}, 025 (2000)
\hepth{0008073}.

\bibitem{ggkrw}
S.~S.~Gubser, S.~Gukov, I.~R.~Klebanov, M.~Rangamani and E.~Witten,
``The Hagedorn transition in non-commutative open string theory,''
hep-th/0009140.

\bibitem{gkp}
S.~Gukov, I.~R.~Klebanov and A.~M.~Polyakov,
``Dynamics of (n,1) strings,''
Phys.\ Lett.\ {\bf B423}, 64 (1998),
\hepth{9711112}.

\bibitem{kleb00}
I.~R.~Klebanov, ``1+1 Dimensional NCOS and its $U(N)$ Gauge Theory
Dual,'' Talk at Strings 2000,
University of Michigan (July 13, 2000),
{\tt
http://feynman.physics.lsa.umich.edu/cgi-bin/s2ktalk.cgi?klebanov}~.

\bibitem{mp}
J.~Maharana and S.~S.~Pal,
``Noncommutative open string, D-brane and duality,''
Phys.\ Lett.\  {\bf B488}, 410 (2000),
\hepth{0005113}.

\bibitem{cw}
G.~Chen and Y.~Wu,
``Comments on noncommutative open string theory: V-duality and  holography,''
\hepth{0006013}.

\bibitem{lrs1}
J.~X.~Lu, S.~Roy and H.~Singh,
``((F,D1),D3) bound state, S-duality and noncommutative open string /  
Yang-Mills theory,''
JHEP {\bf 0009}, 020 (2000),
\hepth{0006193}.

\bibitem{rs-j}
J.~G.~Russo and M.~M.~Sheikh-Jabbari,
``On noncommutative open string theories,''
JHEP {\bf 0007}, 052 (2000),
\hepth{0006202}.

\bibitem{rvu}
S.~Rey and R.~von Unge,
``S-duality, noncritical open string and noncommutative gauge theory,''
\hepth{0007089}.

\bibitem{co}
R.~Cai and N.~Ohta,
``(F1, D1, D3) bound state, its scaling limits and SL(2,Z) duality,''
\hepth{0007106}.

\bibitem{lrs2}
J.~X.~Lu, S.~Roy and H.~Singh,
``SL(2,Z) duality and 4-dimensional noncommutative theories,''
\hepth{0007168}.

\bibitem{gremm}
M.~Gremm,
``Compactified NCOS and duality,''
hep-th/0009095.

\bibitem{bfss}
T.~Banks, W.~Fischler, S.~H.~Shenker and L.~Susskind,
``M theory as a matrix model: A conjecture,''
Phys.\ Rev.\ {\bf D55}, 5112 (1997),
\hepth{9610043}.

\bibitem{susskind}
L.~Susskind,
``Another conjecture about M(atrix) theory,''
\hepth{9704080}.

\bibitem{seiberg}
N.~Seiberg,
``Why is the matrix model correct?,''
Phys.\ Rev.\ Lett.\ {\bf 79}, 3577 (1997),
\hepth{9710009}.

\bibitem{sen}
A.~Sen,
``D0 branes on $T^{n}$ and matrix theory,''
Adv.\ Theor.\ Math.\ Phys.\ {\bf 2}, 51 (1998),
\hepth{9709220}.

\bibitem{bbss2}
E.~Bergshoeff, D.~S.~Berman, J.~P.~van der Schaar and P.~Sundell,
``Critical fields on the M5-brane and noncommutative open strings,''
\hepth{0006112}.

\bibitem{bbss1}
E.~Bergshoeff, D.~S.~Berman, J.~P.~van der Schaar and P.~Sundell,
``A noncommutative M-theory five-brane,''
\hepth{0005026}.

\bibitem{harmark2}
T.~Harmark,
``Open branes in space-time non-commutative little string theory,''
\hepth{0007147}.

\bibitem{kt}
T.~Kawano and S.~Terashima,
``S-duality from OM-theory,''
\hepth{0006225}.


\bibitem{ck}
C.~G.~Callan and I.~R.~Klebanov,
``D-Brane Boundary State Dynamics,''
Nucl.\ Phys.\ {\bf B465}, 473 (1996),
\hepth{9511173}.

\bibitem{verlinde}
H.~Verlinde,
``A matrix string interpretation of the large N loop equation,''
\hepth{9705029}.

\bibitem{motl}
L.~Motl,
``Proposals on nonperturbative superstring interactions,''
\hepth{9701025}.

\bibitem{bs}
T.~Banks and N.~Seiberg,
``Strings from matrices,''
Nucl.\ Phys.\ {\bf B497}, 41 (1997),
\hepth{9702187}.

\bibitem{dvv}
R.~Dijkgraaf, E.~Verlinde and H.~Verlinde,
``Matrix string theory,''
Nucl.\ Phys.\ {\bf B500}, 43 (1997),
\hepth{9703030}.

\bibitem{malda}
J.~Maldacena, ``The Large $N$ Limit of Superconformal Field Theories
and Supergravity,'' Adv.~Theor. Math.~Phys. {\bf 2} (1998) 231,
{\tt hep-th/9711200}.

\bibitem{polchinski}
J. Polchinski, {\it String Theory},
Cambridge University Press (1998), Vol. 1.

\bibitem{bgl}
V.~Balasubramanian, R.~Gopakumar and F.~Larsen,
``Gauge theory, geometry and the large N limit,''
Nucl.\ Phys.\ {\bf B526}, 415 (1998),
\hepth{9712077}.

\bibitem{rindep}
A.~G\"uijosa,
``Is physics in the infinite momentum frame independent of the
compactification radius?,''
Nucl.\ Phys.\ {\bf B533}, 406 (1998),
\hepth{9804034}.

\bibitem{hp}
S.~Hellerman and J.~Polchinski,
``Compactification in the lightlike limit,''
Phys.\ Rev.\ {\bf D59}, 125002 (1999),
\hepth{9711037}.

\bibitem{bilal}
A.~Bilal,
``A comment on compactification of M-theory on an
(almost) light-like circle,''
Nucl.\ Phys.\ {\bf B521}, 202 (1998)
[hep-th/9801047].

\bibitem{bilal2}
A.~Bilal,
``DLCQ of M-theory as the light-like limit,''
Phys.\ Lett.\ {\bf B435}, 312 (1998),
\hepth{9805070}.

\bibitem{uy}
S.~Uehara and S.~Yamada,
``On the DLCQ as a light-like limit in string theory,''
\hepth{0008146}.

\bibitem{gpr}
A.~Giveon, M.~Porrati and E.~Rabinovici,
``Target space duality in string theory,''
Phys.\ Rept.\  {\bf 244}, 77 (1994),
\hepth{9401139}.

\bibitem{weinberg}
S.~Weinberg,
``Dynamics At Infinite Momentum,''
Phys.\ Rev.\  {\bf 150}, 1313 (1966).

\bibitem{ks}
J.~Kogut and L.~Susskind,
``The Parton Picture Of Elementary Particles,''
Phys.\ Rept.\  {\bf 8}, 75 (1973).

\bibitem{my}
T.~Maskawa and K.~Yamawaki,
``The Problem Of P+ = O Mode In The Null Plane 
Field Theory And Dirac's Method Of Quantization,''
Prog.\ Theor.\ Phys.\  {\bf 56}, 270 (1976).

\bibitem{pb}
H.~C.~Pauli and S.~J.~Brodsky,
``Discretized Light Cone Quantization: Solution 
To A Field Theory In One Space One Time Dimensions,''
Phys.\ Rev.\  {\bf D32}, 2001 (1985).

\bibitem{harvey}
J.~A.~Harvey,
``Spin dependence of D0-brane interactions,''
Nucl.\ Phys.\ Proc.\ Suppl.\ {\bf 68}, 113 (1998),
\hepth{9706039}.

\bibitem{wati}
W.~I.~Taylor,
``D-brane field theory on compact spaces,''
Phys.\ Lett.\ {\bf B394}, 283 (1997),
\hepth{9611042}.

\bibitem{lst}
N.~Seiberg,
``New theories in six dimensions and matrix
description of M-theory on $\mathbf{T}^{5}$ and
$\mathbf{T}^5/\mathbf{Z}_{2}$,''
Phys.\ Lett.\ {\bf B408}, 98 (1997),
\hepth{9705221}.

\bibitem{rozali}
M.~Rozali,
``Matrix theory and U-duality in seven dimensions,''
Phys.\ Lett.\  {\bf B400}, 260 (1997),
\hepth{9702136}.

\bibitem{brs}
M.~Berkooz, M.~Rozali and N.~Seiberg,
``Matrix Description of M-theory on $\mathbf{T}^{4}$ and $\mathbf{T}^{5}$,''
Phys.\ Lett.\  {\bf B408}, 105 (1997),
\hepth{9704089}.

\bibitem{hyun}
S.~Hyun,
``U-duality between NCOS theory and matrix theory,''
\hepth{0008213}.


\end{thebibliography}
\end{document}